\documentclass[12pt,a4paper,div=11]{scrartcl}
\usepackage[T1]{fontenc}
\usepackage[utf8]{inputenc}
\usepackage[english]{babel}
\usepackage{amsmath,amsfonts,amssymb,amsthm}	
\usepackage{graphicx}
\usepackage{rotating}	
\usepackage{pdflscape}
\usepackage{subcaption}
\usepackage{multicol,multirow,booktabs,array,longtable}	
\usepackage[flushleft]{threeparttable}
\usepackage{setspace}
\usepackage{xr}
\usepackage{url}
\usepackage{natbib}		
\usepackage{dcolumn}
\usepackage{hyperref} 
\usepackage{color}
\usepackage{lmodern}
\setkomafont{caption}{\normalsize} 
\usepackage{ulem}
\usepackage{cancel}

\DeclareMathOperator{\sgn}{sgn}
\newcommand{\sym}[1]{\rlap{#1}} 
\usepackage{siunitx}
\def\yyy{%
	\bgroup\uccode`\~\expandafter`\string-%
	\uppercase{\egroup\edef~{\noexpand\text{\llap{\textendash}\relax}}}%
	\mathcode\expandafter`\string-"8000 }
\def\xxxl#1{%
	\bgroup\uccode`\~\expandafter`\string#1%
	\uppercase{\egroup\edef~{\noexpand\text{\noexpand\llap{\string#1}}}}%
	\mathcode\expandafter`\string#1"8000 }
\def\xxxr#1{%
	\bgroup\uccode`\~\expandafter`\string#1%
	\uppercase{\egroup\edef~{\noexpand\text{\noexpand\rlap{\string#1}}}}%
	\mathcode\expandafter`\string#1"8000 }

\makeatletter
\edef\originalbmathcode{%
    \noexpand\mathchardef\noexpand\@tempa\the\mathcode`\(\relax}
\def\resetMathstrut@{%
  \setbox\z@\hbox{%
    \originalbmathcode
    \def\@tempb##1"##2##3{\the\textfont"##3\char"}%
    \expandafter\@tempb\meaning\@tempa \relax
  }%
  \ht\Mathstrutbox@\ht\z@ \dp\Mathstrutbox@\dp\z@
}
\makeatother

\usepackage{ulem}
\usepackage{xcolor}

\begin{document}
\title{Decisions and Performance Under Bounded Rationality: A Computational Benchmarking Approach\thanks{
Uwe Sunde gratefully acknowledges financial support of the Deutsche Forschungsgemeinschaft (CRC TRR 190 Rationality and Competition, project number 280092119). Anthony Strittmatter gratefully acknowledges the hospitality of CESifo.\newline
Contact: Anthony Strittmatter anthony.strittmatter@ensae.fr, Uwe Sunde uwe.sunde@lmu.de (corresponding author), Dainis Zegners zegners@rsm.nl
}}

\author{
\normalsize \textsc{Dainis Zegners} \\
{\small Rotterdam}\\
{\small School of Management,}\\
{\small Erasmus University}\\
\and
\normalsize \textsc{Uwe Sunde}\\
{\small LMU Munich} \\
{\small CEPR, Ifo, IZA}
\and
\normalsize \textsc{Anthony Strittmatter} \\
{\small CREST-ENSAE}\\
{\small CESifo}}

\date{}
\maketitle

\vspace{-1cm}
\begin{abstract}
This paper presents a novel approach to analyze human decision-making that involves comparing the behavior of professional chess players relative to a computational benchmark of cognitively bounded rationality. This benchmark is constructed using algorithms of modern chess engines and allows investigating behavior at the level of individual move-by-move observations, thus representing a natural benchmark for computationally bounded optimization. The analysis delivers novel insights by isolating deviations from this benchmark of bounded rationality as well as their causes and consequences for performance. The findings document the existence of several distinct dimensions of behavioral deviations, which are related to asymmetric positional evaluation in terms of losses and gains, time pressure, fatigue, and complexity. The results also document that deviations from the benchmark do not necessarily entail worse performance. Faster decisions are associated with more frequent deviations from the benchmark, yet they are also associated with better performance. The findings are consistent with an important influence of intuition and experience, thereby shedding new light on the recent debate about computational rationality in cognitive processes.

\begin{description}
\item JEL-classification: D01, D9, C7, C8 \smallskip
\item Keywords: Cognitively Bounded Rationality, Benchmark Computing, Artificial Intelligence, Decision Quality, Decision Time
\end{description}
\end{abstract}

\thispagestyle{empty} \setcounter{page}{0} \newpage

\normalsize \renewcommand{\baselinestretch}{1.5}\normalsize


\begin{quote}
``I don't believe in psychology. I believe in good moves.'' \hspace{0.3cm} (Bobby Fischer)
\end{quote}

\section{Introduction}\label{sec: intro}

Research in behavioral economics and decision sciences has identified a plethora of different, partly distinct and partly interacting, behavioral deviations from the predictions of models of rational decision making. These deviations are typically associated with cognitive limitations related to computational complexity, capacities for memory, and related factors such as emotional stress in combination with time pressure. This observation has led to repeated requests to develop and apply models of bounded rationality as a more adequate description of human behavior than perfect rationality, especially in complex decision environments \citep{Simon1979,Simon1982,aumann1997,Camerer1998}.

Nevertheless, the assumption of perfect rationality still prevails as the conventional starting point to study human behavior in economics, not only in theoretical but also in applied work. 
One of the reasons for this is the lack of consensus about an appropriate alternative benchmark for boundedly rational behavior.
In psychology and cognitive science, there is an ongoing debate about the appropriate behavioral model of decision-making in complex environments and about the role of optimization in decision-making. Work along the \textit{heuristics and biases} approach \citep{kahneman1996} mainly views deviations from rationality as errors and hence detrimental for decision performance. In contrast, work in the tradition of \textit{fast and frugal heuristics} maintains that decisions based on heuristics are the result of algorithmic reasoning that involves rules about search, stopping, and decision-making, and can lead to advantageous decisions \citep{gigerenzer1996,gigerenzer_goldstein1996}.
Another strand of work emphasizes the role of experience and intuition for the capability of making adequate decisions in complex, often stressful real-world situations where the assessment of various decision alternatives is often not possible \citep{Klein1993,kahnemanklein2009}. Theoretical work on bounded rationality has typically proposed specific, axiomatic models of bounded rationality in the context of complex choices \citep[see, e.g., ][]{lipman1991,lipman1999, gabaix2006, alaoui2016}, but no dominant benchmark for assessing behavior has emerged. This has led to a discrepancy between the development of modeling tools for decision making and their applicability in the business or policy context \citep[see, e.g.][in the context of dynamic programming]{Rust:2019}, and raised calls for more systematic investigations of behavior while relaxing the assumption of perfect rationality \citep{Iskhakov/etal:2020}.

Empirical work has typically established deviations from rationality by comparing actual behavior against a theoretical benchmark of rationality, often in simplistic, unrealistic, or abstract settings that are unfamiliar to the decision makers. Consistent with the view of heuristics leading to advantageous solutions of decision problems, recent work in cognitive psychology has emphasized evolutionary factors that shape decision making and heuristics \citep{fawcett2014}. Research in computer science has long argued for an application of concepts of bounded optimization that explicitly accounts for constraints related to cognitive capacity and time requirements, instead of applying a rational benchmark \citep{russel1995}. More recently, these insights have been incorporated in conceptual models of computational rationality \citep{lewis2014} or resource-rational analysis \citep{lieder2020}. The main difficulty in this context has been the same that has plagued the conceptualization of bounded rationality, namely the development of ``a meaningful formal definition of rationality in a situation in which calculation and analysis themselves are costly and/or limited'' \citep[p. 12]{aumann1997}.

This paper contributes an analysis of human decision-making by developing a conceptualization of cognitively bounded rationality in a complex real-world setting that is based on a computational benchmark. This benchmark is based on the exclusive notion of bounded optimality as consequence of cognitive constraints. Consistency with, or deviations from, this benchmark thus shed light on the nature of decision making in complex environments. By isolating the role of behavioral factors that have previously been associated with behavioral ``biases'' or heuristics, this approach helps reconciling research on cognitive biases with the view that deviations from a benchmark of cognitively bounded rationality might not necessarily entail negative consequences for performance if deviations are based on intuition and experience that are the result of adaptation of the limitations of depth of reasoning in the decision problem. While related research has typically proposed models that focus on specific features of boundedly rational behavior, our approach starts from a concrete, natural computational benchmark for behavior under cognitive constraints without imposing a specific behavioral model. By comparing decisions in a complex environment against this benchmark, the design of our analysis also allows us to investigate the performance consequences of bounded rationality that deviates from the benchmark of cognitively bounded rationality.

Concretely, we study the decisions of professional chess players who participate in tournaments with high stakes and incentives to win a game. The context of chess is ideally suited for the purpose of this study. Chess provides a clean and transparent decision environment that allows observing individual choices in a sequential game of perfect information. Despite the conceptually simple structure of the decision problem that lends itself to an application of standard dynamic programming techniques, the complexity of the problem rules out perfectly rational behavior in most situations: Even the best existing chess engines are unable to determine the rationally optimal move in the large majority of chess positions.\footnote{Exceptions to this rule are positions during the endgame with a very limited number of pieces on the board, or positions entailing a forced check-mate or forced move repetition leading to a draw within the computational horizon of the chess engine.} Hence, by construction any observed behavior is rooted in bounded rationality. At the same time, professional chess players are commonly seen as the prototypes of ``rational'' and experienced individuals who are familiar with forward-looking strategic behavior.

Our methodology makes use of artificial intelligence embodied in modern chess engines. The strategies suggested by a chess engine that resembles the strength of play of human players constitute a relevant, transparent, and insightful benchmark for bounded rationality in light of a decision task for which the determination of the rational strategy involves an unsurmountable degree of complexity. Hence, this setting allows us to analyze decisions and the corresponding performance consequences relative to a benchmark of bounded rationality stipulated by the objective algorithm of a chess engine that is subject to comparable cognitive and computational limitations. The information about decisions at extremely high resolution and accuracy -- at the level of an individual move -- can then be used to isolate the exact circumstances and psychological factors that lead to deviations from this benchmark of bounded rationality and thereby shed light on behavioral aspects in boundedly rational behavior. In particular, the detailed information about the evaluation of a given configuration, the time left for decision making, complexity, and the time used for a given move provides a unique possibility to decompose different candidates for behavioral factors based on variation within the same person and game.


The analysis compares the actual moves of human decision makers to the best conceivable move in the respective configuration. This best conceivable move is determined by a ``super chess engine'' whose performance and computational capacity exceeds that of the best humans by far. We also replicate each configuration observed in a large data set of chess games and determine the decision of a ``restricted chess engine'' that computes the best move under the assumption of mutual best response, but that is comparable in terms of playing strength to human players and simulates play against another chess engine of similar strength. Like the human decisions, these replicated decisions of the restricted chess engine are then evaluated in comparison to the moves suggested by the super chess engine. The analysis of the difference-in-differences makes it possible to identify behavioral deviations of professional chess players from the objective benchmark of bounded rationality constituted by the restricted engine.
This setting can also shed light on the consequences of these deviations for performance using within-person variation at the level of individual moves. This allows investigating not only whether humans behave differently, but whether they perform better, compared to the boundedly rational benchmark provided by chess engines, and under which circumstances.

The results provide new insights into human decision making and document systematic deviations of human behavior from the benchmark of bounded computational rationality in several dimensions. These deviations can be related to different behavioral factors that have been discussed in the previous literature that compared behavior to the rational benchmark. While behavioral deviations from rationality are usually associated with suboptimal performance, this connotation often rests on a priori reasoning or value judgments as it is typically even harder to identify the consequences for performance of deviations from the benchmark than the deviations themselves. The approach developed here allows investigating not only whether humans behave differently than a computational benchmark of bounded optimality, but also whether they perform better than stipulated by this benchmark. In particular, we find systematic deviations of human behavior in relation to the current standing reflected in terms of an advantage or disadvantage. Being in a better position induces deviations from the objective benchmark that are associated with worse performance than stipulated by the benchmark, while being in a worse position is associated with more deviations that are associated with better performance. A smaller remaining time budget in the game leads to more frequent deviations and worse performance, suggestive of the detrimental effects of time pressure. We also find evidence for the role of fatigue over the course of a game, which reduces the likelihood of deviations with better performance. An intensification of cognitive limitations in the context of particularly complex configurations leads to more frequent deviations from the rational benchmark, but not to a systematic deterioration in performance. When investigating the mechanisms, we find no systematic differences in the causes and consequences of behavioral deviations from the benchmark between weaker and stronger players. Strategic interactions or psychological factors, as reflected by the remaining time of the opponent,
seem of limited importance.

To shed light on the ongoing debate in psychology regarding bounded optimality and computational rationality as adequate representations of decision making, we also explore a core implication, namely the allocation of decision times. In particular, if the factors that lead to deviations from the computational benchmark are related to intuitive reasoning, potentially reflecting heuristics, they should systematically interfere with the time spent on deliberating a particular decision. An analysis of decision times reveals that being substantially ahead or behind entails faster decisions, as does time pressure and fatigue. Greater complexity induces longer deliberation times, consistent with decision makers devoting more effort to solve a more demanding decision problem. When considering the relation to deviations from the computational benchmark, and their performance implications, we find that faster decisions are associated with more frequent deviations from the rational benchmark, but at the same time are associated with better performance. This evidence is suggestive of a superior intuitive assessment of particular configurations, which is presumably related to memory, experience or other factors relevant for the decision-making process.

\paragraph{Contribution to the Literature.} The results of this paper contribute to a substantial literature
that has used chess as the prime example of how to think about and model strategic behavior. Analyses of optimal behavior in chess laid the grounds of game theory, with early proofs of the existence of winning strategies by \citet{Zermelo:1913} and equilibrium by \citet{Neumann:1928}; see \citet{Schwalbe/Walker:2001} for an informative overview. Chess players have a long history as subjects of studies in psychology, starting with the work of \cite{deGroot1946,deGroot1978}. \citet{chase1973} and \citet{simon1973} contain early discussions of theories of cognition derived from the study of chess players.
Work in psychology on expert performance regularly uses chess players as subjects of study \citep{ericsson2006,moxley2012}. The view of professional chess players as the prototypes of rational decision makers led several empirical or experimental tests of rational behavior in economics to focus on chess players as subjects of interest. Examples include experiments with chess players to investigate the empirical relevance of subgame perfection and backward induction \citep{Palacios/Volij:2009,Levitt/etal:2011}, rational learning in repeated games \citep{Gerdes/etal:2010}, or emotions and psychological factors \citep{Diaz/Palacios:2016}.
Data from chess tournaments have also been used to analyze various other research questions. These include, in particular,
gender differences in patience \citep{Gerdes/etal:2011}, gender effects in competitiveness \citep{Backus/etal:2016,sousa16}, gender and attractiveness \citep{Dreber/etal:2013}, self-selection and productivity in  tournaments \citep{Bertoni/etal:2015, Linnemmer/Visser:2016}, consequences of political ideology \citep{Frank/Krabel:2013}, collusion \citep{Moul/Nye:2009}, cheating \citep{barnes2015, Regan2015} and indoor air quality \citep{Kuenn/etal:2019}. To our knowledge, this is the first study to analyze move-by-move behavior of chess players relative to a benchmark of bounded rationality provided by a chess engine of comparable chess strength to human players.

Recent work has used chess engines to assess the relative performance and strength of chess players in different time periods \citep{guid2011,Alliot:2017}. \citet{Anderson/Green:2018} use chess data at the player-game level to investigate the role of personal peak performance in the past in terms of ratings, as reference points for performance. \citet{Strittmatter/Sunde/Zegners:2020a} use data on the player-game level over the past 125 years to estimate the cognitive performance over the life cycle and its dynamics over time and across cohorts. \citet{Anderson/etal:2016} develop a prediction model of errors based on difficulty, skills and time restricting attention to positions with 5 or less pieces on the board (so-called endgame table bases) to compare the moves of humans to a benchmark of perfect play. 
Recent work by \cite{Anderson2020} develops a neural network to predict moves by human chess players. In contrast to the existing work in this literature, which typically analyzes human performance at the game level and uses a chess engine that vastly outperforms human chess players to benchmark behavior, the methodology developed here allows us to identify deviations from a benchmark of cognitively bounded decision making by comparing behavior to an engine of comparable strength on a move-by-move basis, as well as the performance implications of these deviations along the lines of a differences-in-differences setting.

Our computational approach to a benchmark of bounded rationality complements earlier work that proposed specific models of bounded rationality in the context of complex choices \citep[see, e.g., ][]{lipman1991,lipman1999, gabaix2006, alaoui2016}. Instead of providing evidence for behavior being in line with a specific model, our approach is to explore the drivers of deviations from the benchmark and their performance consequences, in line with the research program on computational rationality \citep{gershman2015}. For a long time, chess players were hypothesized to heavily rely on intuition and expertise in their decision making instead of applying bounded optimization consistent with computational rationality \citep[e.g.][]{simon1973,kahnemanklein2009,moxley2012}. However, to our knowledge there exists no clear evidence on the implications of deviations from bounded rationality for performance. By documenting that deviations from bounded rationality do not necessarily imply worse performance but can even lead to better performance than the benchmark, our evidence also contributes to recent theoretical work on foundations of behavior. Our results provide evidence that is consistent with the predictions of recent theoretical work that has considered the optimal speed and accuracy of decisions in settings in which the relative evaluations of decision alternatives are unknown; the results of this work show that decision accuracy may actually decrease with longer decision time \citep{Fudenberg/etal:2018}. Likewise, the result that deviations from the rational benchmark can be associated with better performance is consistent with predictions of models of focusing and selective memory \citep{Gennaioli/Shleifer:2010,Bordalo/Gennaioli/Shleifer:2020} or case-based decision theory \citep{Sahm/Weizsaecker:2016}. Finally, our results are consistent with conjectures in psychology of the importance of intuitive expertise and recognition-primed decision making complex environments \citep{Shanteau1992,Schultetus/Charness:1999,kahnemanklein2009}.

Most contributions in behavioral economics have focused on documenting a behavioral deviation from the rational benchmark in one particular dimension in an abstract setting and theoretical work on bounded rationality has typically focused on one particular aspect. This paper analyzes field evidence that allows us to isolate behavioral factors that lead to deviations from a natural benchmark of behavior in a setting that involves bounded rationality. The comparison of behavior to an objective benchmark in terms of the quality of a given move relative to the best possible move in a given configuration allows us to explore the empirical relevance of various behavioral factors within a single and comparable research design, as well as their implications for performance. Our results thereby complement findings of the detrimental effects of time pressure on the quality of decision making \citep{Kocher/Sutter:2006} and relate to findings of heterogeneous effects of time pressure in loss and gain domains \citep{Kocher/etal:2013}. Our findings also contribute to the literature that has emphasized the role of choking under pressure \citep{Baumeister:1985,Krumer:2017,Dohmen:2008,Genakos/etal:2015} or limited attention \citep{Foellmi/etal:2016} among professionals. The heterogeneity in the results for deviations from rational behavior depending on the current positional standing in the game in terms of advantage or disadvantage is also reminiscent of findings of reference dependence \citep{Bartling/etal:2015} and observations from risk taking in tournaments \citep{Cabral2003, Genakos/Pagliero:2012}. Likewise, the results add to the literature investigating the role of complexity and cognitive load for individual performance \citep{Deck/Jahedi:2015} and on the relationship between cognitive limitations and behavioral biases \citep{Oechssler/etal:2009,Stracke/etal:2017}.


The remainder of the paper is structured as follows. Section \ref{sec: data} contains a description of the data collection and measurement. Section \ref{sec: empirical strategy} develops the empirical approach. Section \ref{sec: results} presents the empirical results. Section \ref{sec: conclusion} concludes.

\section{Data and Measurement}\label{sec: data}
\subsection{Data from Professional Chess Players}

In the terminology of game theory, chess is a two-person, sequential, zero-sum game with perfect information and alternating moves, for which an optimal strategy exists.\footnote{See \citet{Schwalbe/Walker:2001} for details and a discussion of the historical background.}  
The data used in the empirical analysis have been collected from an internet platform that broadcasts all professional over-the-board chess tournaments (\url{www.chess24.com}) and contains detailed information for more than 100,000 moves from around 2,000 games that were played in 97 single round-robin tournaments during the years 2014-2017. All games were played at regular time controls that allocate a time budget of a minimum of 2 hours thinking time to each player to conclude the game.\footnote{According to the regulations by the International Chess Federation FIDE, for a game to be rated each player must have a minimum of 120 minutes, assuming the game lasts 60 moves per player. The standard time control regime suggested by the International Chess Federation FIDE is 90 minutes per player per game plus 30 seconds added to each player's time budget for each move played; additional 30 minutes are added to each player's time budget after each player has played 40 moves (see \url{https://handbook.fide.com}, last accessed May 12, 2020). 
Tournaments that are not officially organized by FIDE use slight variations of the official FIDE time control regime.} Appendix Table \ref{tab:tournaments} provides an overview of the tournaments contained in the data set.
The data contain detailed information about the players, including their performance statistics in terms of their ELO number.\footnote{The ELO number constitutes a method for calculating the relative playing strength of players (invented by the Hungarian mathematician Arpad Elo). The ELO number increases or decreases depending on the outcome of games between rated players. After every game, the winning player takes points from the losing player, while the total number of points remains fixed. According to international conventions, an ELO number of at least 2,500 is a requirement for being awarded the title of an international grandmaster (this requirement has to be fulfilled once during the career, but does not have to be maintained to keep the title, see \texttt{https://handbook.fide.com/chapter/B01Regulations2017}, last accessed April 20, 2020).}
We restrict our baseline analyses to games between professional players with an ELO number of at least 2,500 at the time of the game. Appendix Table \ref{tab:summary_games} shows summary statistics on the game level.


In addition to the remaining time budget and time consumed for each move, the move-by-move data comprises information about the exact configuration of pieces on the board. We use this information to compute an evaluation of this configuration in terms of the relative standing of each player, an evaluation of the complexity of the configuration, and an evaluation of move quality, as explained in more detail below. For the computation of performance, we exclude the first fifteen moves of each player in a game from our analysis. These are usually so-called ``book moves'', which are studied intensively by players in the preparation of the game and are typically the result of routine openings.

\subsection{Measuring Performance in Chess}

To construct a benchmark for behavior, we make use of a chess engine, \textsc{Stockfish 8}, which is an open-source program that computes the best possible move for a given configuration of pieces on the chessboard. This engine is considered to be one of the best available programs. The version we use has an estimated ELO rating of approximately 3150 points (in comparison, the incumbent World Champion Magnus Carlsen had an ELO rating of 2872 points in January 2020 according to the official rating list by the International Chess Federation FIDE).\footnote{We limit \textsc{Stockfish 8} to a search depth of 21 moves to economize on computing costs. The unconstrained version of \textsc{Stockfish 8} has an ELO of approximately 3300 points (\url{http://ccrl.chessdom.com}). Based on an approximation by \cite{ferreira2013}, the ELO strength of \textsc{Stockfish 8} with search depth of 21 corresponds to approximately 3150 points.
}
This engine can be restricted, such that the strength of play of the engine corresponds closer to the strength of play of human players. To construct an objective benchmark of behavior for human players, we use engines with different strengths of play in the analysis as described in detail below.

\begin{figure}[]
\begin{center}
\includegraphics[width=0.7\textwidth]{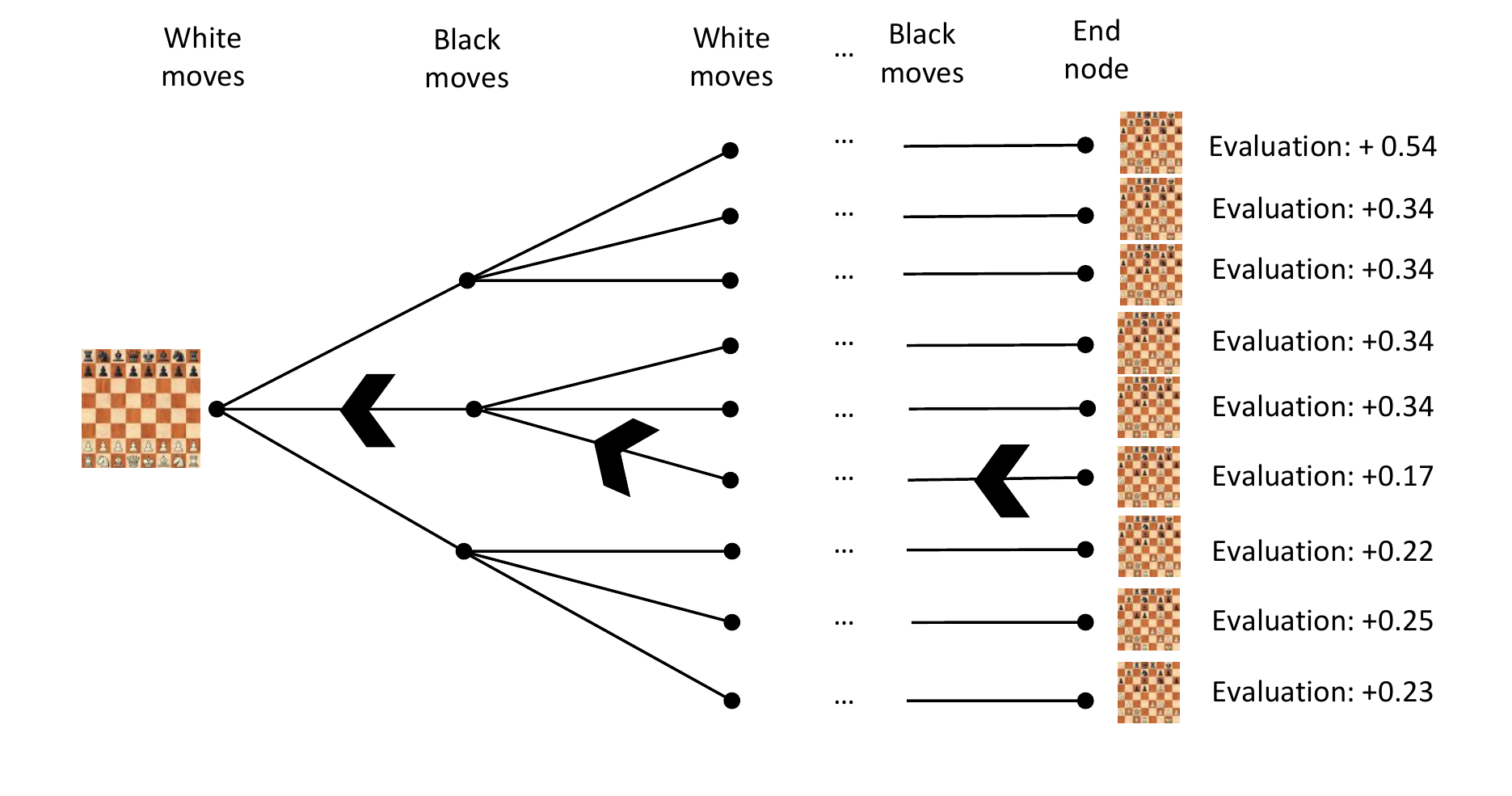} \\
  \caption{Backward Induction in Chess Engines}\label{fig: game tree}
\vspace{0.2cm}
\parbox{15cm}{
 \footnotesize \emph{Note:} Illustration of the decision algorithm built into a chess engine. For a given search depth (number of moves until the end node is reached), the engine calculates evaluations of different alternative moves under the assumption of mutually best response and determines the move that delivers the highest evaluation on the end node. The positions at the end nodes are evaluated using a human-curated evaluation function that considers factors of the chess positions such as number of pawns and pieces on the board, pawn structure, mobility of pieces and king safety.}
\end{center}
\end{figure}

An engine behaves exactly as stipulated by standard game-theoretic considerations subject to intellectual (computational) constraints. For each configuration, the engine creates a game-tree for all possible moves by white and black for a pre-specified length of $n$ moves ahead, the so-called search depth. Then, the configurations at the respective end-nodes are evaluated in terms of pieces left on the board, safety of the king, mobility of pieces, pawn-structure and so on. Based on this evaluation, the engine then determines the best move using backward induction under the assumption of mutually best responses.\footnote{Modern chess engines are almost exclusively based on domain-specific algorithmic heuristics that were developed specifically to search the sequential game-tree arising from a given configuration. Current chess engines use an enhanced version of the min-max algorithm that disregards branches of the search tree that have already been found to be dominated. This reduces the search-space without impacting the final choice of the best move by the engine (\url{https://www.chessprogramming.org/Alpha-Beta}, last visited March 17, 2020). Modern engines like \textsc{Stockfish 8} calculate approximately 10-100 million nodes per second on standard personal computing hardware. Only very recently more general machine learning techniques in the form of neural networks have been embodied in chess engines such as Google's non-public AlphaZero  \citep{silver2018}.} Figure \ref{fig: game tree} illustrates the decision algorithm solved by the engine. This delivers a clean and transparent benchmark to evaluate human behavior.

\begin{figure}[]
\begin{center}
\includegraphics[width=0.9\textwidth]{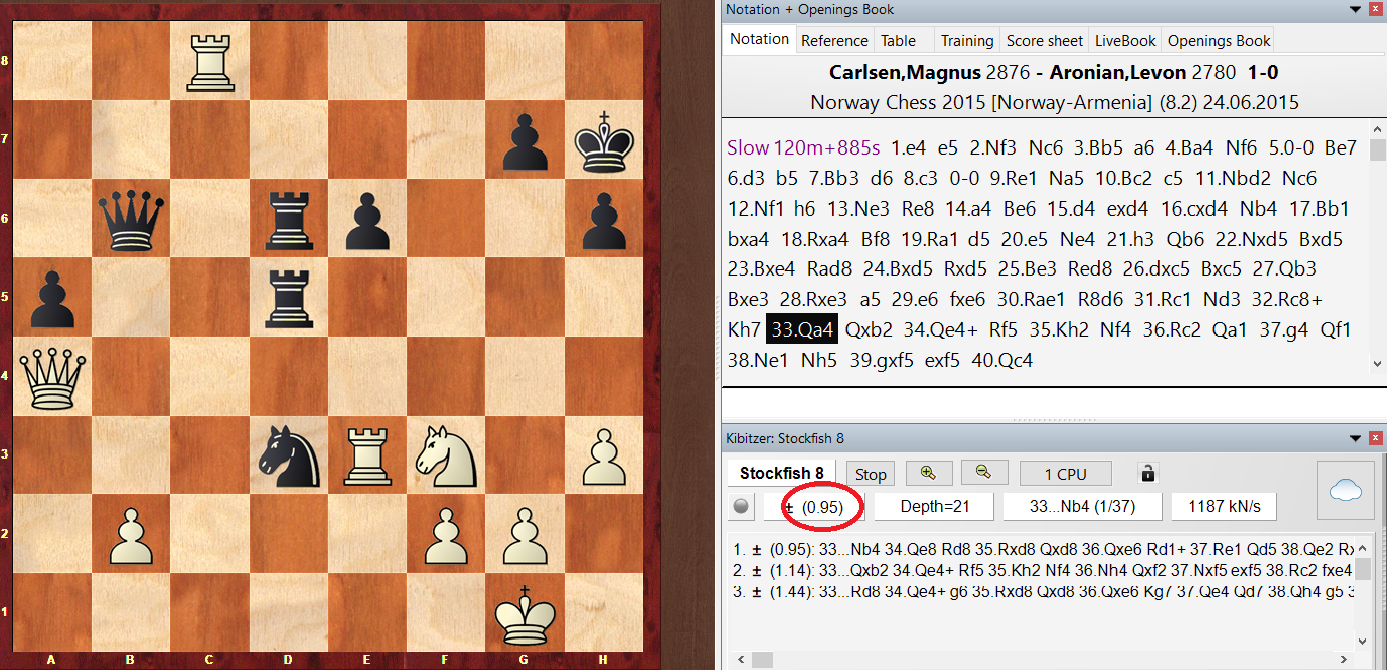}
\caption{Computation of Performance Measures: An Example}\label{fig: evaluation example}
\end{center}
\footnotesize{\emph{Note:} The engine evaluates the configuration shown on the board as +0.95 (i.e., an advantage for white of almost one pawn unit) if Black plays Knight to b4 as the next move. Instead, Black played Queen takes b2, which the engine judges as a slight mistake, with the consequence of an evaluation of +1.14 for White after this move. Hence, the quality of Black's move is computed as -0.19, i.e., Black played a move that resulted in the loss of 0.19 pawn units compared to the evaluation resulting after the move suggested by the engine. In this example, the engine needed 3.87 seconds to reach a search-depth of 21 moves, which corresponds to the measure of complexity of the configuration.}
\end{figure}

We use a chess engine with superior performance compared to any human player (the ``super engine'') to compute three measures that are central to the empirical analysis. First, the engine delivers a measure of the relative standing for a given configuration of pieces on the board, which reflects an evaluation of the current position of a player and represents a proxy of the winning odds. The evaluation of the current position is the result of the engine computing, for each configuration observed in the data set, the best continuation. The relative standing is measured in so-called pawn units, where one unit approximates the advantage of possessing one more pawn.\footnote{This measure is relative and indicates an advantage for the player with white pieces for positive numbers, and for the player with black pieces for negative numbers. For example, if the evaluation is -1.00 pawn units, black is better ``as if one pawn up.''}
Second, we compute a measure of performance, in terms of the quality of play of a given player, by comparing the actual move made by the respective player to the best move suggested by the chess engine. This move is not necessarily the absolutely best move that is possibly conceivable but on average the move suggested by the engine is better than conceivable by any human player. In the data, relative performance can be measured by a binary indicator of whether a player makes the optimal move (or one of the optimal moves in case of several moves with equal winning odds) as suggested by the chess engine in a given configuration. Alternatively, one can construct a measure of the quality of a move by computing the distance of a player's move (in terms of the conventional metric of pawn units) from the best move identified by the chess engine.\footnote{Concretely, we configure the engine to compute the corresponding evaluations for the six moves that it evaluates as best in a given configuration. Increasing the number of moves that are evaluated further comes at a prohibitively large computational cost. If the actual move played is one of these six moves, the performance is calculated as the difference in evaluation between the best and the actual move played. If the move played is not among the six best moves, we compute performance as the difference in the evaluation right before and right after the respective move of the player.}
Third, we use the engine to compute, for each observed configuration, a measure of complexity of the configuration. The more complex the configuration, the longer the engine needs to search the game-tree. The time consumed by the engine to compute the best strategy for the next $n$ moves ahead can therefore be used as a measure of the complexity of a given configuration.\footnote{As baseline measure of complexity, we use the computation time needed by the super chess engine to reach a search-depth of 21 moves. Alternatively, we use the number of branches (nodes) of the game-tree that the engine has to calculate to reach a search-depth of 21 as a measure of the branching factor and thus complexity of a given configuration. The (unreported) results are qualitatively similar and available upon request.}
Figure \ref{fig: evaluation example} contains a concrete illustration of how these measures are computed.

Appendix Table \ref{tab: descriptive statistics} documents the descriptive statistics of the move-by-move data used in the analysis.

\section{Empirical Strategy}\label{sec: empirical strategy}

\subsection{Conceptual Approach}

We denote by $P_{ic}$ the performance of a move by human player $i$ in a given configuration $c$ of pieces relative to the benchmark suggested by the super chess engine. This suggestion is based on the backward induction algorithm described before and constitutes a first natural benchmark for studying behavioral patterns. The relative performance measure $P_{ic}$ is not sufficient for isolating behavioral deviations from a benchmark of bounded rationality, however, because the measure does not account for the fact that the super chess engine is superior to human players in terms of strength of play. Thus, the objective human evaluation of a given configuration and the resulting optimal human move might differ systematically from the suggested optimal move of the super chess engine as a result of differences in the cognitive limitations that give rise to bounded rationality.

To address this issue, the empirical strategy applies a difference-in-differences logic that compares the performance of humans to the performance of an equally strong engine where, in both cases, performance is measured relative to the best possible move based on the assessment of the super chess engine. To construct such a directly comparable benchmark of cognitively bounded rationality with similar playing strength as humans, we replicate each decision problem faced by humans (for each configuration $c$ observed in our data set) using a restricted chess engine that is calibrated to have approximately the same strength of play as the humans observed in the data set. This implies that, for each observed configuration $c$, we construct a benchmark performance measure for a strength of play comparable to that of the human players (with ELO numbers between 2500 and 2880) relative to the best possible move suggested by the super chess engine.
In particular, we restrict the chess engine to a fixed depth of reasoning (in terms of search depth) that is constant across chess positions in our data.\footnote{We restrict \textsc{Stockfish 8} to a search depth of 12 moves,
which corresponds to a play strength equivalent to an ELO of around 2700 when comparing performance differences between human players and the restricted engine (see Appendix Figure \ref{fig:Human_and_Restricted_Chess_Engine}).} This allows us to identify deviations from this benchmark that are due to humans adapting their reasoning and behavior to the nature of the decision problem.

By construction, the restricted engine plays best response strategies such that the move played by the restricted chess engine only depends on objective, move-specific characteristics related to the configuration on the board, but not on subjective player-specific or game-history-related factors. Deviations of performance of this restricted engine from the best possible move suggested by the super engine can thus be due only to cognitive limitations, but not due to behavioral deviations that reflect bounded rationality in humans. The relative performance of the restricted chess engine thus delivers a valid performance benchmark of cognitively bounded rationality against which the performance of humans can be compared. Notice that a plain comparison of moves between humans and the restricted chess engine would not be sufficient, because it would not be possible to evaluate the direction -- and thus the performance consequences -- of these differences. This is only achieved by the comparison to the best possible move suggested by the super chess engine.

\subsection{Parameters of Interest}

To illustrate the identification strategy, let the potential relative performance under the computational benchmark of cognitively bounded rationality
in configuration $c$ be denoted by $P_{c}^*$. This is a potential variable that is not observed; we only observe the realized relative performance $P_{ic}$ in the data, which might differ from $P_{c}^*$ because of behavioral deviations. Performance differences due to deviations from this benchmark are defined by $D_{ic}  = P_{ic}- P_{c}^*$, where $D_{ic}=0$ implies no deviation from the benchmark. Notice that also $D_{ic}$ is a potential variable that is unobserved. For ease of notation, define the dummy variable $D_{ic}^E= \mathbb{I}\{D_{ic}\neq0\} $, with $\mathbb{I}\{\cdot\}$ being the indicator function, as an indicator of any deviation from the benchmark.

The goal of the empirical analysis is to identify subjective (psychological) factors $X_{ic}$ that can be associated with deviations from the benchmark of cognitively bounded rationality. The conditional expectation of behavioral deviations is
\begin{equation}\label{iter}
E[D_{ic}|X_{ic}=x] =  E[D_{ic}|D_{ic}^E=1,X_{ic}=x]\cdot p(x),
\end{equation}
where $p(x) = Pr(D_{ic}^E=1|X_{ic}=x)$ denotes the conditional probability of a behavioral deviation from the benchmark. The right hand side of equation \eqref{iter} makes use of the discrete law of iterated expectations and the fact that $E[D_{ic}|D_{ic}^E=0,X_{ic}=x]=0$. The marginal effects of subjective factors $X_{ic}$ on deviations can be decomposed into effects along the extensive and intensive margin conditional on deviation, since
\begin{equation} \label{decom}
\underbrace{\frac{\partial E[D_{ic}|X_{ic}]}{\partial X_{ic}}}_{\mbox{\scriptsize Total Effect}} =  E[D_{ic}|D_{ic}^E=1,X_{ic}]\cdot \underbrace{\frac{\partial p(X_{ic})}{\partial X_{ic}}}_{\mbox{\scriptsize Extensive Margin}} + \underbrace{\frac{\partial E[D_{ic}|D_{ic}^E =1,X_{ic}]}{\partial X_{ic}}}_{\mbox{\scriptsize Intensive Margin}}\cdot p(x) .
\end{equation}

Deviations from the benchmark in terms of performance differences as reflected by $D_{ic}$ are sensitive to the respective metric in which they are measured (e.g., pawn units). The extensive margin effects have the advantage to not depend on the particular metric of $D_{ic}$.
To explore the consequences of deviations from the benchmark of cognitively bounded rationality at the extensive margin, we denote positive deviations, i.e., deviations from the benchmark that are associated with better performance than the benchmark, by $D_{ic}^P= \mathbb{I}\{D_{ic}>0\} $. Likewise, negative deviations, i.e., behavioral deviations from the benchmark that are related to worse performance are denoted by $D_{ic}^N= \mathbb{I}\{D_{ic}< 0\} $. Furthermore, we denote the conditional probability of a behavioral deviation from the benchmark that implies better performance by $p_{P}(x)=Pr(D_{ic}^P|X_{ic}=x)$ and the conditional probability of a behavioral deviation that implies worse performance by $p_{N}(x)=Pr(D_{ic}^N|X_{ic}=x)$, with $p(x) = p_{P}(x)+ p_{N}(x)$.
The partial effects along the extensive margin can then be decomposed into partial effects on the probabilities of behavioral deviations associated with positive and negative consequences for performance,
\begin{equation*}
\frac{\partial p(X_{ic})}{\partial X_{ic}} = \underbrace{\frac{\partial p_{P}(X_{ic})}{\partial X_{ic}}}_{\mbox{\scriptsize Positive Consequences}} + \underbrace{\frac{\partial p_{N}(X_{ic})}{\partial X_{ic}}}_{\mbox{\scriptsize Negative Consequences}}\:.\\
\end{equation*}


\subsection{Identification}

We now sketch an identification strategy that allows identifying the parameters of interest. Let $P_{c}^{r}$ denote the relative performance in configuration $c$ by the restricted chess engine (with strength of play similar to that of humans) in comparison to the performance under an optimal move suggested by the super engine.
Furthermore, we denote the difference between the relative performance of humans and the restricted chess engine by $\Delta_{ic} = P_{ic}- P_{c}^{r}$.

As a first step in the identification of the effects of subjective factors $X_{ic}$, we focus on the effects along the extensive margin. For this purpose, we construct a binary measure of whether the relative performance of a human player differs from the restricted chess engine, $\Delta_{ic}^{E} = \mathbb{I}\{\Delta_{ic}\neq0\} $. This binary measure represents the observable analogue to $D_{ic}$ and the identification of the effects along the extensive margin relies on the assumption that
\begin{equation}\label{eq:identification assumption}
E[\Delta_{ic}^{E}- D_{ic}^E|X_{ic}=x] = 0 \:.
\end{equation}
This assumption is fundamentally not testable, because deviations from the performance benchmark, $D_{ic}^E$, are unobservable. The assumption implies that the conditional probability that human players deviate from the restricted engine is equal to the conditional probability that human players deviate from the performance benchmark, which is a natural assumption in our setting. In the data, 60\% of all moves exhibit the same relative performances for humans and the restricted chess engine. Accordingly, there is a mass point in the distribution of $\Delta_{ic}$, which is otherwise a continuously distributed variable.
Notice that the exact calibration of the strength of the restricted engine might influence the results by influencing the empirical measure of $\Delta^E$. However, as discussed below, extensive robustness tests show that the results are insensitive to variations in the calibration of the restricted engine.


Under assumption \eqref{eq:identification assumption}, the  marginal effect of a subjective (psychological) factor $X_{ic}$ on the probability of observing a deviation from the benchmark is given by\footnote{This follows from
\begin{equation*}
E[\Delta^E_{ic}|X_{ic}=x] = E[\Delta_{ic}^{E}-D_{ic}^E |X_{ic}=x] + E[D_{ic}^E|X_{ic}=x]  = E[\Delta_{ic}^{E}-D_{ic}^E |X_{ic}=x] + p(x),
\end{equation*}
and noting that $E[\Delta_{ic}^{E}-D_{ic}^E |X_{ic}=x]=0$  under assumption \eqref{eq:identification assumption}.}
\begin{equation*}
 \frac{\partial E[\Delta_{ic}^{E}|X_{ic}]}{\partial X_{ic}} = \frac{\partial p(X_{ic})}{\partial X_{ic}} .
\end{equation*}
This effect corresponds to an effect along the extensive margin and contains no information about the performance implications of this deviation.

Next, consider the marginal effects on the probability of behavioral deviations that imply better or worse performance than the benchmark, respectively. For this purpose, define $\Delta_{ic}^{P} = \mathbb{I}\{\Delta_{ic}> 0\} $ and $\Delta_{ic}^{N} = \mathbb{I}\{\Delta_{ic}< 0\} $. Under the assumption
$E[\Delta_{ic}^{P}-D_{ic}^P|X_{ic}=x] = 0$,
the marginal effects of factors $X$ on $\Delta^P$ correspond to marginal changes in the probability of behavioral deviations with better performance,
\begin{equation*}
\frac{\partial E[\Delta_{ic}^{P}|X_{ic}]}{\partial X_{ic}} = \frac{\partial p_{P}(X_{ic})}{\partial X_{ic}} \:.
\end{equation*}
Similarly, under the assumption
$E[\Delta_{ic}^{N}-D_{ic}^N|X_{ic}=x] = 0$,
the marginal effects on changes in the probability of behavioral deviations with worse performance are given by
\begin{equation*}
\frac{\partial E[\Delta_{ic}^{N}|X_{ic}]}{\partial X_{ic}} = \frac{\partial p_{N}(X_{ic})}{\partial X_{ic}} \:.
\end{equation*}
Using the categorical measure $\Delta_{ic}^{C} =  \sgn(\Delta_{ic}) \cdot(1-\mathbb{I}\{\Delta_{ic}=0\})$, these insights can be combined to obtain the net effect on the probability of deviations with positive and negative performance consequences,
\begin{equation*}
  \frac{\partial E[\Delta_{ic}^{C}|X_{ic}]}{\partial X_{ic}} = \frac{\partial p_{P}(X_{ic})}{\partial X_{ic}} - \frac{\partial p_{N}(X_{ic})}{\partial X_{ic}}\:,\\
\end{equation*}
provided that the previous assumptions
hold.\footnote{Note that the assumptions $E[\Delta_{ic}^{P}-D_{ic}^P|X_{ic}=x] = 0$ and  $E[\Delta_{ic}^{N}-D_{ic}^N|X_{ic}=x] = 0$ together are somewhat stronger than assumption \eqref{eq:identification assumption}.}

%


Finally, reconsider the total marginal effect of the subjective factors $X_{ic}$ as described in equation \eqref{decom}.
Using the performance measure $\Delta_{ic}$, this effect can be identified under the
assumption $E[P_{c}^*- P_{c}^{r}|X_{ic}=x] = 0$, such that
\begin{equation*}
\frac{\partial E[\Delta_{ic}|X_{ic}]}{\partial X_{ic}} =  \frac{\partial E[D_{ic}|X_{ic}]}{\partial X_{ic}} \:,
\end{equation*}
which is a combination of the effects along the extensive and intensive margin.\footnote{In particular,
\begin{align*}
E[\Delta_{ic}|X_{ic}=x] =& E[P_{ic}-P_{c}^*|X_{ic}=x] + E[P_{c}^*- P_{c}^{r}|X_{ic}=x] \\
 =& E[D_{ic}|D_{ic}^E=0,X_{ic}=x] \cdot \left(1-p(x)\right) + E[D_{ic}|D_{ic}^E=1,X_{ic}=x]\cdot p(x)\\
  =&  E[D_{ic}|D_{ic}^E=1,X_{ic}=x]\cdot p(x),
\end{align*}
which follows from applying the discrete law of iterated expectations similarly as in \eqref{iter}.}
The intensive margin effects of the subjective factors $X_{ic}$ conditional on deviation are identified
under the assumption that \eqref{eq:identification assumption} and $E[P_{c}^*- P_{c}^{r}|X_{ic}=x] = 0$
both hold.\footnote{In particular,
\begin{equation*}
 E[\Delta_{ic}|\Delta_{ic}^{E}=1,X_{ic}=x]  =\frac{ E[\Delta_{ic}|X_{ic}=x]}{Pr(\Delta_{ic}^E=1|X_{ic}=x)} =\frac{ E[D_{ic}|X_{ic}=x]}{p(x)}= E[D_{ic}|D_{ic}^E=1,X_{ic}=x]\:.
\end{equation*}
The first and last equalities follow from the discrete law of iterative expectations. The second equality holds under $E[P_{c}^*- P_{c}^{r}|X_{ic}=x] = 0$ (numerator) and assumption \eqref{eq:identification assumption} (denominator).} Then,
\begin{equation*}
\frac{\partial E[\Delta_{ic}|\Delta_{ic}^{E}=1,X_{ic}]}{\partial X_{ic}} =  \frac{\partial E[D_{ic}|D_{ic}^E=1,X_{ic}]}{\partial X_{ic}} \:.
\end{equation*}
The interpretation of the intensive margin effects conditional on deviation is problematic, however, because the subjective factors affect the performance consequences of deviations and the probability of observing a deviation at the same time, thus giving rise to a sample selection problem \citep[see, e.g.,][]{heck79}.

\subsection{Estimation}

In practice, we use move-by-move data with an observation for the positional configuration of pieces on the board $c$ faced by individual player $i$ in game $g$. The estimation model is then given by
\begin{equation}\label{eq:individual performance}
  \Delta_{gic}' = X_{gic} \beta + \phi_{ig} + u_{gic}\:,
\end{equation}
with the error term $u_{gic}$. 
$\Delta_{gic}'$ denotes the different performance measures ($\Delta^E_{gic}$, $\Delta^P_{gic}$, $\Delta^N_{gic}$, $\Delta^C_{gic}$, $\Delta_{gic}$) described above. All specifications include interacted player-game fixed effects $\phi_{ig}$ (where subscript $ig$ indicates the player-game-level) to account for systematic variation in style of play, environmental factors related to the game, or strategic aspects related to particular pairings. Inference is based on game-level clustered standard errors to account for interdependencies in the performances of both players.

The parameter vector $\beta$ represents the partial effects of different subjective (psychological) factors $X_{gic}$, $\beta = \partial E[\Delta_{gic}'|X_{gic}]/\partial X_{gic}$. In view of earlier work, we primarily focus on four subjective (psychological) factors $X_{gic}$ that might affect behavioral deviations from the objective benchmark: emotions and preferences related to the current standing (being in a better or worse position), time pressure (remaining time budget), fatigue (number of moves played by each player before the current move), and complexity (related to cognitive limitations).

\section{Behavioral Deviations from the Benchmark}\label{sec: results}

\subsection{Main Results}

Table \ref{tab:human_deviate} contains the results of multivariate regression analyses of the empirical model in equation \eqref{eq:individual performance} for different dependent variables. Column (1) shows coefficient estimates for regressions with the binary measure of any deviation from the benchmark, $\Delta^E$, as dependent variable. Compared to an approximately balanced positional standing, human players are more likely to deviate from the benchmark when being in a better position relative to their opponent. In contrast, they are not more likely to deviate in a worse position. The results for remaining time reveal a positive but only marginally significant effect on the probability to deviate from the benchmark. This suggests that players deviate more often from the benchmark if they have more time available, rather than under greater time pressure. Contradicting intuition regarding a potential influence of fatigue, the probability to deviate from the benchmark is smaller later on in the game. Greater complexity of the configuration is associated with a higher probability to deviate from the benchmark.

\begin{table}[t]
\caption{Behavioral Deviations from Cognitively Bounded Rationality\label{tab:human_deviate}}
\footnotesize
{
\def\sym#1{\ifmmode^{#1}\else\(^{#1}\)\fi}
\begin{tabular*}{1\hsize}{@{\hskip\tabcolsep\extracolsep{-10pt}}l*{4}{D{.}{.}{-1}}}
\toprule
 &\multicolumn{4}{c}{Dependent Variable:}\\
                    &\multicolumn{1}{c}{$\Delta^E$ (binary)}&\multicolumn{1}{c}{$\Delta^P$ (binary)}&\multicolumn{1}{c}{$\Delta^N$ (binary)}&\multicolumn{1}{c}{$\Delta^C$ (categ.)}\\\cmidrule(lr){2-2}\cmidrule(lr){3-3}\cmidrule(lr){4-4}\cmidrule(lr){5-5}
                    &\multicolumn{1}{c}{(1)}         &\multicolumn{1}{c}{(2)}         &\multicolumn{1}{c}{(3)}         &\multicolumn{1}{c}{(4)}         \\
\midrule
\emph{Current Position}&                     &                     &                     &                     \\
Better position (>0.5 pawnunits)&      0.0796\sym{***}&      0.0006         &      0.0790\sym{***}&     -0.0784\sym{***}\\
                    &    (0.0060)         &    (0.0043)         &    (0.0048)         &    (0.0069)         \\
Worse position (<-0.5 pawnunits)&      0.0057         &      0.0306\sym{***}&     -0.0248\sym{***}&      0.0554\sym{***}\\
                    &    (0.0061)         &    (0.0046)         &    (0.0050)         &    (0.0075)         \\
\bottomrule
\end{tabular*}
}
\footnotesize{\emph{Note:} {OLS estimates. Evaluations of performance are based on the \textsc{Stockfish 8} chess engine (super engine and restricted engine). The variable \textit{Num. previous moves} is calculated as the number of previous moves per player. Standard errors are clustered on the game level. $^{*}$: $p<0.1$, $^{**}$: $p<0.05$, $^{***}$: $p<0.01$. }}
\end{table}

Columns (2) and (3) present the results for the binary measures of deviations from the benchmark that also contain information about the direction in terms of the associated consequences for performance, $\Delta^P$ and $\Delta^N$. Here, a somewhat richer picture emerges. Whereas being in a better position is not associated with human players deviating in a way that their performance is better than the benchmark (Column (2)), the effect on deviations that imply worse performance than the benchmark is positive and significant (Column (3)). A possible explanation for this finding is that human players might decide to play sub-optimal moves that are associated with lower risk or complexity, but also worse performance, when in a better position. The opposite picture emerges when players are in a worse position. In this case, humans are more likely to make deviations that imply better performance than the benchmark (Column (2)), but are less likely to make deviations that imply worse performance than the benchmark (Column (3)). This finding is consistent with stronger incentives for higher performance, for instance due to loss aversion relative to a reference point of a balanced position. As a consequence, humans might become less focused or more adventurous when they are in a better position, but they excel when they are in a worse position.

The picture also becomes richer regarding the influence of time pressure. More remaining time increases the likelihood of deviations with better performance than the benchmark (Column (2)), whereas the likelihood of deviations with worse performance declines (Column (3)). This provides evidence that deviations with worse performance become more frequent with less remaining time, consistent with the hypothesis of choking under time pressure. These opposite effects for deviations with different consequences for performance also explain why remaining time only has a weakly positive effect on the probability of any deviation (Column (1)). Likewise, a clearer picture emerges regarding fatigue, proxied by the number of moves that have already been played during a game. In particular, later in the game, deviations from the benchmark that are associated with better performance become less frequent, whereas there is no significant effect on the likelihood of deviations that are associated with worse performance. Finally, the hypothesis that complexity affects deviations from the benchmark is supported by significant effects on deviations with both, higher and lower performance than the benchmark. This is consistent with the conjecture that it is harder for human players to determine the optimal continuation in more complex settings. However, this only results in higher variability of decision quality, but not necessarily in worse average performance.

Column (4) of Table \ref{tab:human_deviate} presents results for the categorical measure $\Delta^C$ as dependent variable. This measure allows making inference on the difference between the effects obtained for $\Delta^P$ and $\Delta^N$. In particular, the estimates confirm the findings that players in a better position are more likely to exhibit worse performance than the benchmark, whereas players that are in a worse position are more likely to deviate with better performance than the benchmark. Also the result for time pressure becomes more pronounced, indicating that less remaining time is associated with more frequent deviations and worse performance. Fatigue continues to imply more frequent deviations from the benchmark with performance deteriorating later in the game. Finally, the effect of complexity is significantly negative in the estimates for the categorical variable, but quantitatively small.

Overall, these findings documents that factors like relative position, time pressure, fatigue and complexity lead to deviations from the computational benchmark of cognitively bounded rationality. At the same time, some of these deviations entail better performance than the benchmark, indicating that intuition and experience, possibly even heuristics, provide humans with a capacity of problem solving that can dominate the computational benchmark in terms of performance. These deviations from the benchmark are still consistent with the notion of bounded rationality and computational rationality in the sense of trading off effort, time, complexity, and quality of a move.
These results thereby complement existing work by documenting the empirical prevalence of various behavioral factors related to reference points \citep[see, e.g.][]{Bartling/etal:2015}, time pressure \citep[see, e.g.][]{Kocher/Sutter:2006,Kocher/etal:2013}, psychological pressure \citep[see, e.g.][]{Dohmen:2008}, fatigue, or cognitive load related to complexity \citep[see, e.g.][]{Oechssler/etal:2009,Deck/Jahedi:2015,Stracke/etal:2017}. In contrast to this previous work, we explore the relevance of these different factors within a single framework. More importantly, our analysis is based on the notion of a benchmark of cognitively bounded rationality, which, among other things, allows us to shed light on performance consequences of various behavioral factors. Before we explore the process underlying the observed behavior in more detail in Section \ref{sec:computational rationality}, we document the robustness of the findings and report evidence on effect heterogeneity.

\subsection{Robustness}

\paragraph{Alternative Model Specifications.}
The results are broadly similar when considering specifications without player-game fixed effects (see Appendix Table \ref{tab:baseline_no_FE}). Moreover, controlling for the subjective factors in univariate specifications confirms the robustness of the main findings (Appendix Table \ref{tab:baseline} shows this exemplarily for the categorical variable ($\Delta^C$)). The pattern of the main results remain similar when we exclude moves in positions that the engine evaluates as exactly equal for both players, presumably because the optimal continuation results in a repetition of moves (see Appendix Table \ref{tab:baseline_excluding_position_0}).

\begin{figure}[t]
\begin{center}
\begin{subfigure}{0.45\textwidth}
\includegraphics[width=1\textwidth]{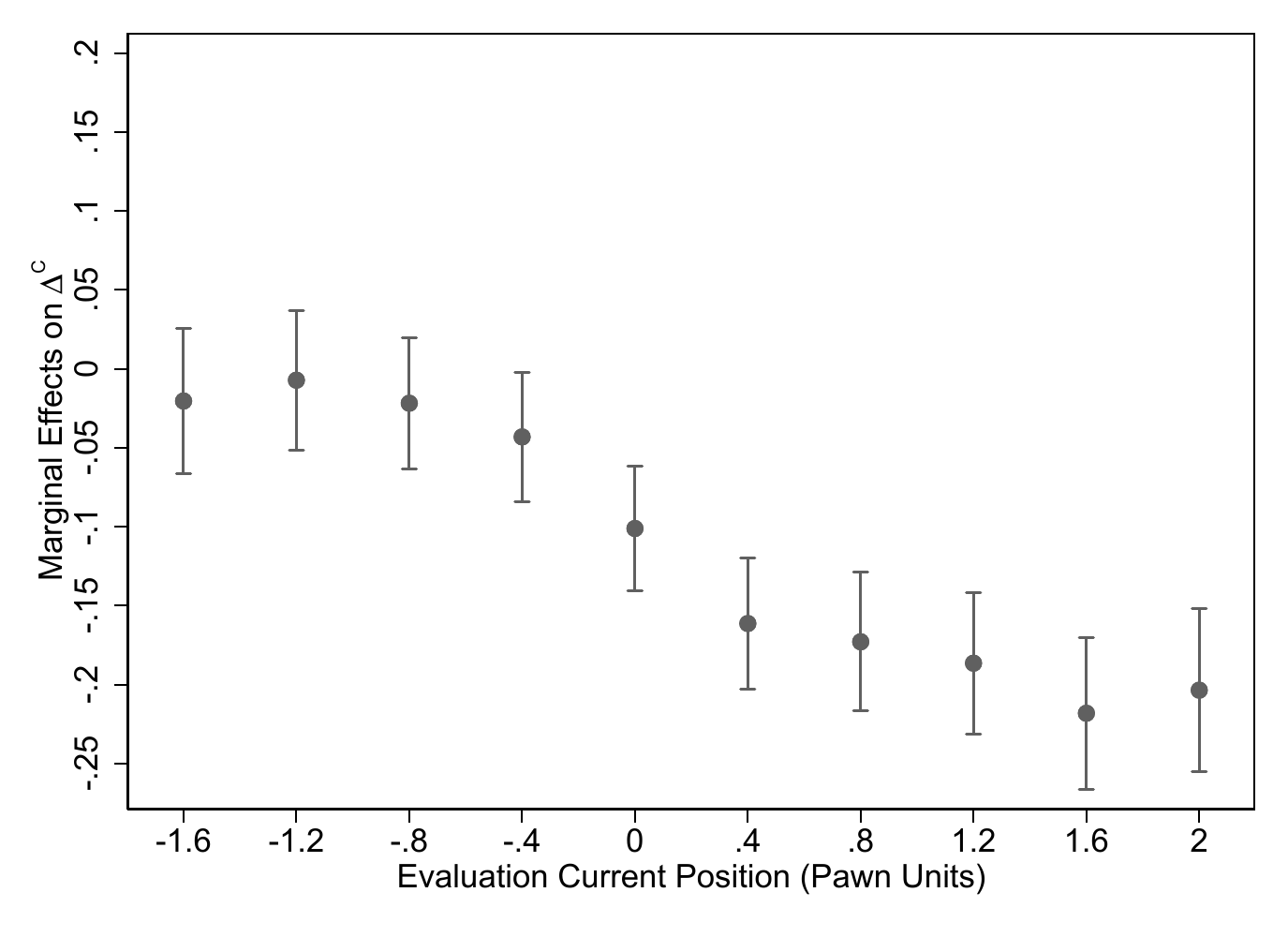}
\caption{Current Position}
\end{subfigure}
\begin{subfigure}{0.45\textwidth}
\includegraphics[width=1\textwidth]{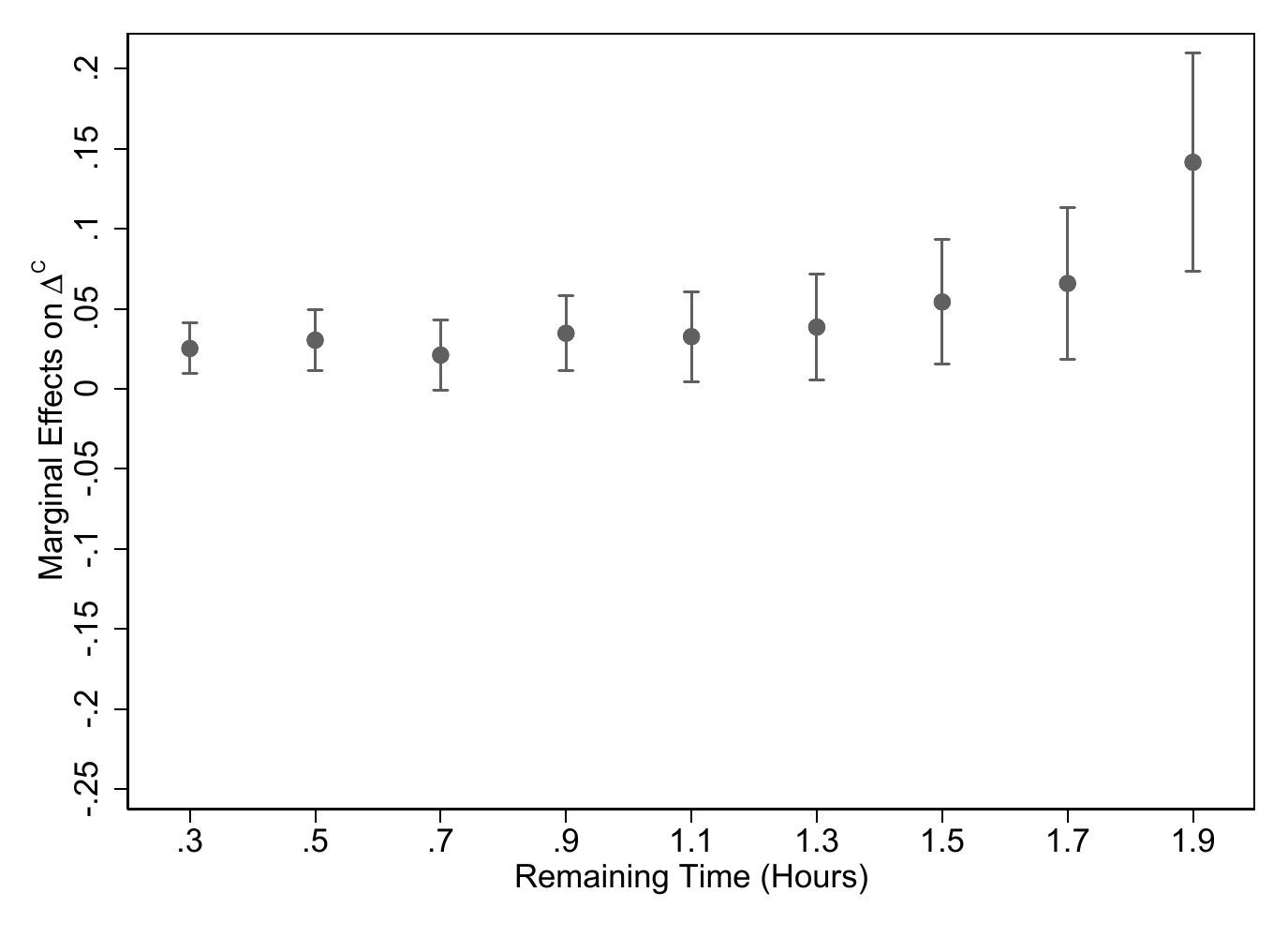}
\caption{Time Pressure}
\end{subfigure} \\
\begin{subfigure}{0.45\textwidth}
\includegraphics[width=1\textwidth]{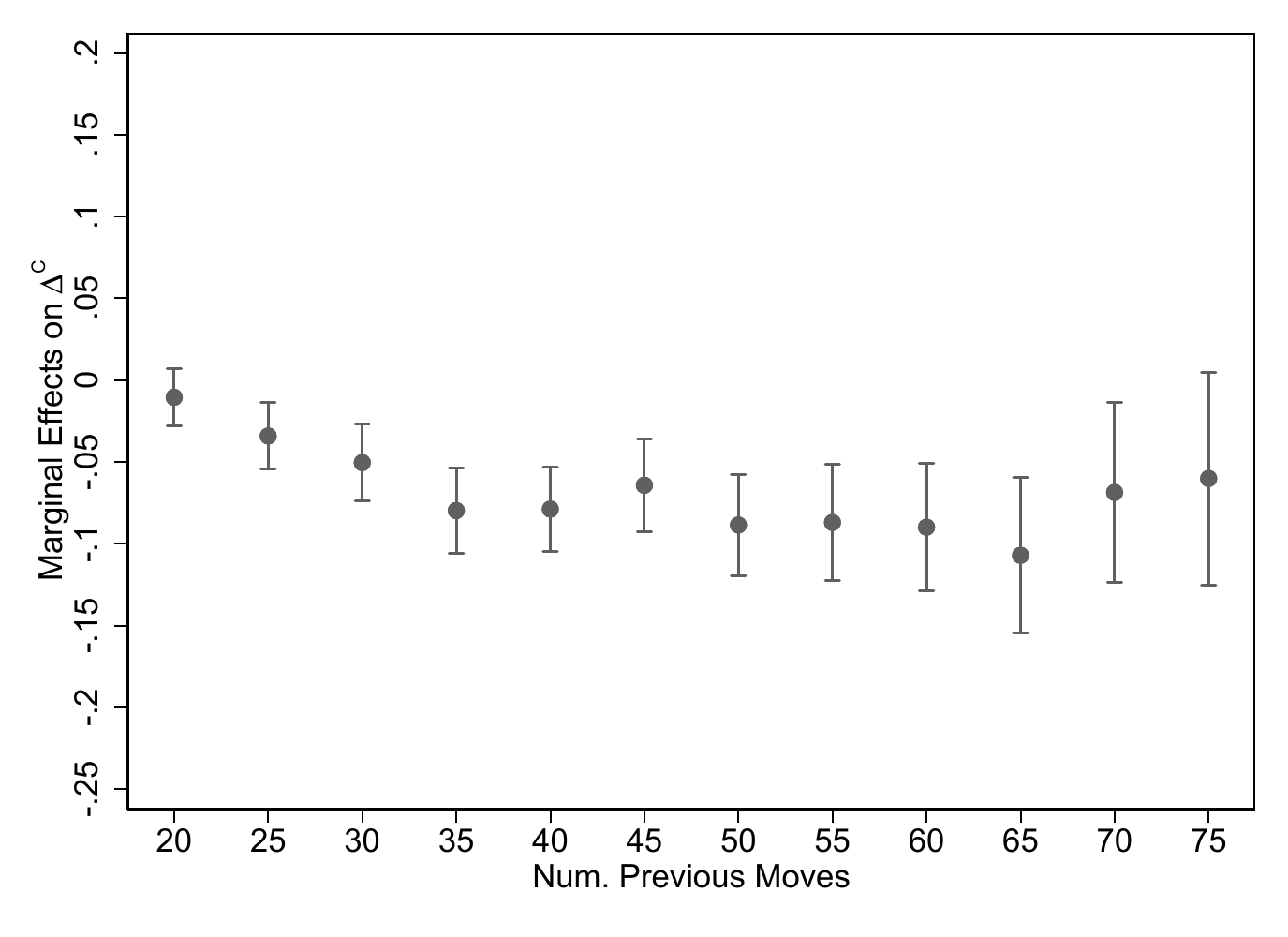}
\caption{Fatigue}
\end{subfigure}
\begin{subfigure}{0.45\textwidth}
\includegraphics[width=1\textwidth]{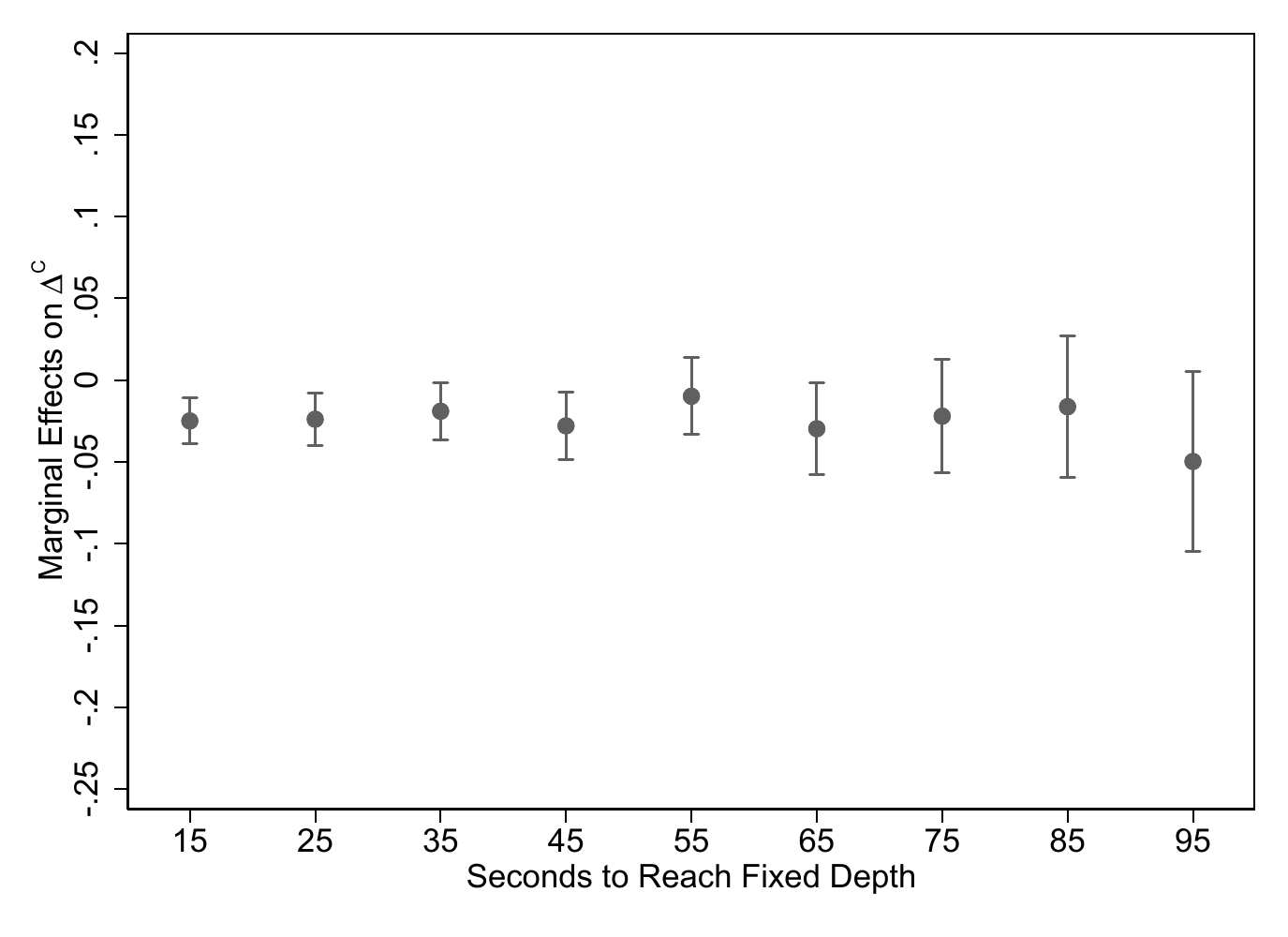}
\caption{Complexity}
\end{subfigure}
\end{center}
\caption{Deviations from Cognitively Bounded Rationality -- Categorical Measure ($\Delta^C$)\label{fig: conditional results}}
\vspace{0.3cm}
\footnotesize{\emph{Note: }{OLS results of more flexible specifications. Dependent variable is $\Delta^C$ and the specification contains player-game fixed effects as in 
the specification in \eqref{eq:individual performance}. Variables depicted on the horizontal axis are split into equal spaced intervals. Dots report point estimates  and whiskers report 95\% confidence intervals. The variable \textit{Num. previous moves} is calculated as the number of previous moves per player.}}
\end{figure}

The estimation results obtained with a more flexible specification of the effect of relative positional standing, allowing for non-linear effects, confirm the main findings and does not reveal evidence for pronounced non-linearities in the effect (see Appendix Table \ref{tab:Baseline_split_better_position_five}). Figure \ref{fig: conditional results} shows results for more flexible specifications of the subjective factors graphically (exemplarily for the dependent variable $\Delta^C$). Figure \ref{fig: conditional results}(a) plots the estimates  from a more flexible specification with respect to current relative standing. The results confirm the main findings of Table \ref{tab:human_deviate}, which reports the results relative to balanced positional standings. Performance is worse for a positive evaluation of the current position compared to a balanced position, but relatively better for negative evaluations. As in the main analysis, the identification of these effects relative to the benchmark of the restricted engine rules out that this finding is driven by mechanical effects such as reversion to the mean.
Figure \ref{fig: conditional results}(b) suggests laxer time budget lead to more frequent deviations with better performance, especially early in the game. Figure \ref{fig: conditional results}(c) and (d) confirm that fatigue and complexity lead to worse performance.

\paragraph{Alternative Measures for the Subjective Factors.} To investigate the robustness of the results, we also replicated the analysis with alternative proxy measures for the various dimensions of behavioral deviations. These include, in particular, relative standing measured in terms of a continuous measure (in pawn units), time pressure in terms of proximity to time control when additional time is added to players' time budget, fatigue as proxied by elapsed time, and complexity in terms of the distance of the second-best move to the first-best move (in terms of pawn units). The results confirm the main results (see Appendix Table \ref{tab:Baseline_five_alternate_measures}).
In comparison to the baseline results, players in a worse position when using a continuous evaluation measure of relative standing are here even significantly less likely to deviate from the benchmark along the extensive margin (for $\Delta^E$), but still exhibit better performance when using the categorical measure $\Delta^C$.


\paragraph{Calibration of Restricted Chess Engine.} Another potential concern regarding the empirical strategy is the calibration of the restricted chess engine. In particular, since identification relies on different assumptions that involve a comparison between human behavior and the computational benchmark of cognitively bounded rationality reflected the restricted engine, the results might be sensitive to the particular calibration as it might induce measurement error in $\Delta_{gic}'$. The empirical specification already accounts for this by including interacted player-game fixed effects that capture potential measurement error that enters at the player-game level, e.g., because a particular player has a systematically higher or lower strength of play than the restricted chess engine. Moreover, the analysis is based on a fairly homogeneous sample of players with ELO ratings between 2,500 and 2,880 points. As discussed above, the results are robust even when the player-game level fixed effects are omitted (see Appendix Table \ref{tab:baseline_no_FE}). Furthermore, measurement error in the response variable does not lead to bias in the coefficient estimates of $\beta$ when it is statistically independent of the regressors, but might increase the variance \citep[see the discussion about classical measurement error in, e.g.,][]{Wool:2010}. Accordingly, statistically independent measurement error may lead to conservative inference.

The most direct evidence for the insensitivity of the results with respect to the calibration of the restricted engine emerges from estimates conducted with subsamples for players with different strength of play. The results from the corresponding robustness checks document that the results are not sensitive to players with different strength of play or the exact specification of the chess engine. In particular, we find that the overall pattern of results is identical when including weaker players (with ELO ratings above 2000 instead of restricting to players with ELO ratings above 2,500), or when restricting to players with ratings between 2,400 and 2,600, or between 2,600 and 2,800 (Appendix Tables \ref{tab:baseline above_2000_elo}, \ref{tab:baseline btw_2400_2600_elo} and \ref{tab:baseline btw_2600_2800_elo}).

\paragraph{Alternative Chess Engine.} To assess the robustness of the results with respect to the particular chess engine, we also replicated the analysis for an alternative engine to construct the benchmark. This engine (\textsc{Komodo}) is considered to have a different playing style than \textsc{Stockfish 8}.\footnote{We use \textsc{Komodo 9}, which is also considered to be among the world's strongest chess engines. \textsc{Komodo}'s playing style is typically referred to as being more positional, focusing more on long-term strategic planning, than that of \textsc{Stockfish}.
According to  \href{https://web.archive.org/web/20190910145801/http://ccrl.chessdom.com/ccrl/4040/cgi/compare_engines.cgi?family=Houdini&print=Rating+list&print=Results+table&print=LOS+table&print=Ponder+hit+table&print=Eval+difference+table&print=Comopp+gamenum+table&print=Overlap+table&print=Score+with+common+opponents}{http://ccrl.chessdom.com} (archived on September 10, 2019) it is estimated to have an ELO of 3235 in its unconstrained version. To replicate moves for the benchmark, we also restrict Komodo to a search depth of 12. 
}
To the extent that the alternative engine exhibits a different playing style, it also potentially introduces different measurement error in the dependent variables $\Delta_{gic}'$ than \textsc{Stockfish 8}, because it uses different computational algorithms. The results obtained with the alternative engine reveal similar patterns as the baseline results (see Appendix Table \ref{tab:human_deviate Komodo}). 

\paragraph{Total Effect/Intensive Margin.} The analysis so far focused on the extensive margin effects. To obtain estimates of the total effects and of the intensive margin effects conditional on deviation as in equation \eqref{decom}, we also estimated the model using a continuous performance measure as dependent variable. In particular, we consider deviations from the benchmark using the measure $\Delta$ in terms of pawn units. Since the distribution of pawn units is substantially skewed and since we consider a semi-continuous variable with a mass point at 0, we construct behavioral deviations from the benchmark in terms of log-modulus transformed performance, $\Delta^L$.\footnote{In particular, we compute $\Delta^L$ as $\Delta^L_{gic} = \sgn(\Delta_{gic})\cdot \ln(|\Delta_{gic}| + 1)$, where $\Delta_{gic}=P_{gic}-P^r_{c}$ is the difference in performance measured in pawn units.}
Recall that identification relies on the assumption that the conditional expectation of the relative performance of humans under objectively optimal behavior is equal to the conditional expectation of the relative performance of the restricted engine, $E[P_{c}^*- P_{c}^{r}|X_{ic}=x] = 0$. This implies a reliance on the particular metric used for measuring performance (here pawn units), in contrast to the identifying assumption for effects along the extensive margin stated in \eqref{eq:identification assumption}. The latter stipulates that the conditional probability of deviations of the relative performance of human players from the benchmark of the restricted chess engine is equivalent to the conditional probability of deviations from the objective performance benchmark, which does not rely on a particular metric. Moreover, the size of the estimated effects depends on the particular metric used, which effectively determines the scope of the intensive margin effect.

Nevertheless, for completeness, we report the estimates of the total effect and the effect along the intensive margin conditional on deviation (see Appendix Table \ref{tab:baseline log modulus} Columns (1) and (2), respectively). In terms of interpretation, the total effect is an alternative measure for the overall performance consequences of behavioral deviations. Comparing the corresponding results to the baseline results for the dependent variable $\Delta^C$ reveals mostly the same patterns for the total effect as for the categorical measure $\Delta^C$ (see Appendix Table \ref{tab:baseline log modulus} Column (1)). The only exception in this pattern refers to the effect of being in a worse position, which exhibits a significantly negative total effect on performance. This effect is quantitatively smaller than the effect for being in a better position but of opposite sign compared to the extensive margin effect of being in a worse position. This suggests that being in a worse position increases the probability of deviations associated with better performance (in terms of $\Delta^C$), but the negative performance effects along the intensive margin associated with worse performance dominate when using the log-modulus transformed performance measure. The other results are qualitatively comparable; complexity has no significant impact on the total effect. The results for the intensive margin effects conditional on deviation are also in line with the findings obtained of the categorical measure $\Delta^C$ (see Appendix Table \ref{tab:baseline log modulus} Column (2)). The exception is again the effect of worse position, which is negative but quantitatively small and only marginally significant. The intensive margin effect for complexity is positive and significant, but also quantitatively small. These patterns are confirmed when using the alternative engine to construct the benchmark (see Columns (3) and (4) of Appendix Table \ref{tab:baseline log modulus}).

In light of the more restrictive identification assumptions, the reliance on a particular performance metric, and the difficult interpretation because of sample selection \citep[see, e.g.,][]{heck79}, we view these findings as reassuring for the overall pattern of results. We conclude that the main insights of the analysis are obtained from the qualitative results along the extensive margin, which have the advantage of a straightforward interpretation and of not relying on a particular performance metric. However, these findings also cast a note of caution regarding the interpretation of various and sometimes diverging findings in the empirical literature on deviations from a rational benchmark, which might not be directly comparable as they result from different outcome measures and thus constitute estimates of effects that are not necessarily fully comparable.

\subsection{Behavioral Heterogeneity}

To shed light on the underlying  behavioral mechanisms, we estimated various alternative specifications that allow for interactions between the factors that lead to behavioral deviations with time pressure, or for heterogeneity in the effects of the subjective factors. Time pressure in terms of less remaining time tends to amplify the probability of any deviations (in terms of $\Delta^E$) associated with better or worse positions, but do not affect the deviations associated with fatigue or complexity. However, time pressure seems not to amplify the consequences of behavioral deviations on performance (in terms of $\Delta^C$), except for complexity where less remaining time is associated with more frequent deviations and even worse performance (Appendix Table \ref{tab:interaction_time_pressure}). These results complement earlier findings for asymmetric effects of time pressure \citep{Kocher/etal:2013}.

A conjecture that has been raised repeatedly in psychology is that stronger players benefit from better intuition \citep{simon1973,moxley2012}. To test this conjecture, we explore whether there is any heterogeneity in the effects of the subjective factors on deviations from the benchmark with respect to player strength, measured by ELO ratings. The results reveal no systematic patterns except that the behavioral deviations associated with time pressure are less pronounced for stronger players (see Appendix Table \ref{tab:effect heterogeneity}).

To study the potential role of reference dependence based on ex-ante odds along the lines of earlier work \citep[e.g.,][]{Bartling/etal:2015} or a potential role of emotional states as in work by \cite{Diaz/Palacios:2016}, we also test for systematic heterogeneity in the performance of players playing with white or black pieces. Playing with white is typically associated with an inherent first-mover advantage at the outset of a game and therefore exhibits significantly higher ex-ante winning odds. Alternatively, we test for heterogeneity across favorites and underdogs as defined by the relative rating of the two players prior to the game in terms of their ELO numbers. However, in our specification with player-game fixed effects, we find no evidence for significant differences in behavioral deviations along these dimensions (see Appendix Table \ref{tab:RefDep_Better_Worse}).

To explore the role of strategic and psychological interactions, we also investigate the influence of the opponent's remaining time or of the time spent on the previous move by the opponent, which reveals no statistically significant interactions between the opponents in terms of an impact on performance (see Appendix Table \ref{tab:time_spent by opponent}).

\section{Evidence for Computational Rationality?\label{sec:computational rationality}}

\subsection{Decision Times}

To shed light on the ongoing debate in psychology and cognitive sciences about whether the appropriate behavioral model of decision-making in complex environments involves optimization, the remaining analysis explores the relation between decisions, performance, and decision times. The notion of computational rationality \citep{gershman2015} stipulates that optimal decisions in complex environments require approximations and, particularly, an efficient use of scarce resources such as time and effort. Consistently, also many heuristics involve rules about search for optimal decisions, as well as rules for optimal stopping and making a decision-making \citep{gigerenzer1996,gigerenzer_goldstein1996}.

We explore the consistency of computational rationality with our evidence in two steps. In this section, we investigate whether the factors that lead to deviations from the computational benchmark are also reflected in decision times. In the next section, we investigate the role of decision times for decisions and their performance.


Table \ref{tab:time_used} presents results for decision time as dependent variable, using an otherwise identical empirical approach as before. The results complement the previous results and are consistent with a model of computational rationality. In particular, the analysis of the determinants of decision time reveals that deliberation is relatively more time consuming when a player is in a better position. The results of the full specification in Column (5) reveals that deliberation is most time consuming in configurations where a player has a clearly better positional standing, while decisions are made faster in the context of worse positions. Presumably, this reflects a greater salience of possible or optimal moves in such situations, but it is also consistent with a greater reliance on intuition and experience for how to proceed. A more constrained time budget in terms of less remaining time at the decision maker's disposal is also associated with less time spent on a move. Hence, under pressure decision makers might be more inclined to rely on their intuition or experience to make a decision. Later in the game, with a greater number of moves played, decisions are made faster. This suggests shorter deliberation as a consequence of fatigue, which indicates a greater reliance on intuition and heuristics as consequence of tighter constraints on cognitive capacities. Finally, more complex situations induce slower decisions. In light of recent findings that longer deliberation times are associated with greater effort and depth of reasoning \citep{alosferrer2020}, this suggests that decision makers need to invest more effort to get to a decision, consistent with a rational use of resources even when decisions have to be made under bounded rationality \citep{lewis2014,gershman2015,lieder2020}.


\begin{table}[t!h]
\begin{center}
\caption{Time Spent on Move as Dependent Variable\label{tab:time_used}}
\footnotesize
{
\def\sym#1{\ifmmode^{#1}\else\(^{#1}\)\fi}
\begin{tabular*}{1\hsize}{@{\hskip\tabcolsep\extracolsep{-24pt}}l*{5}{D{.}{.}{-1}}}
\toprule
 &\multicolumn{5}{c}{Dependent Variable:}\\
                    &\multicolumn{5}{c}{Time spent per move (min.)}                                                               \\\cmidrule(lr){2-6}
                    &\multicolumn{1}{c}{(1)}         &\multicolumn{1}{c}{(2)}         &\multicolumn{1}{c}{(3)}         &\multicolumn{1}{c}{(4)}         &\multicolumn{1}{c}{(5)}         \\
\midrule
\emph{Current Position}&                     &                     &                     &                     &                     \\
Better position (>0.5 pawnunits)&     -0.1857\sym{***}&                     &                     &                     &      0.4894\sym{***}\\
                    &    (0.0599)         &                     &                     &                     &    (0.0506)         \\
Worse position (<-0.5 pawnunits)&     -1.2136\sym{***}&                     &                     &                     &     -0.2344\sym{***}\\
                    &    (0.0618)         &                     &                     &                     &    (0.0470)         \\
\bottomrule
\end{tabular*}
}
\smallskip
\end{center}
\footnotesize{\emph{Note:}{ OLS estimates. The variable \textit{Num. previous moves} is calculated as the number of previous moves per player. Standard errors are clustered on the game level. $^{*}$: $p<0.1$, $^{**}$: $p<0.05$, $^{***}$: $p<0.01$.}}
\end{table}

\subsection{Decision Times and Deviations from the Benchmark}

The results presented in the last section indicate that behavioral deviations from the benchmark do not necessarily imply worse performance. This suggests that human intuition and experience might be an important factor in determining a successful strategy. The findings on decision times complement this interpretation. To explore this issue in relation to the question about computational rationality, we investigate the role of the time players invest in making a decision about a move for decision making in comparison to the computational benchmark. If time allocation is determined by implicit cost-benefit considerations, decision makers spend more time deliberating a move when the gap in the subjective evaluation between two options is relatively small \citep{Chabris/etal:2009}. Moreover, earlier studies found that additional time for deliberation improves performance \citep{moxley2012}. Recent theoretical work has considered the optimal speed and accuracy of decisions in settings in which the relative evaluations of decision alternatives are unknown. This work has shown that \textit{faster} decisions can also imply \textit{better} performance when decision makers already have fairly precise information and the value of further information acquisition is low, or when decision makers face (subjectively) simple problems where information acquisition is fast \citep{Fudenberg/etal:2018}. Our setting allows us to provide new evidence for the relation between decision speed and performance. Under the premise that, for certain configurations, intuition and expert assessment based on experience lead to a fast and precise assessment of the best strategy with little gain from additional deliberation, this gives rise to the hypothesis that faster decisions might be associated with more frequent deviations from the benchmark, but at the same time are not necessarily associated with worse performance. Indeed, if humans deliberately make a decision based on an intuitive assessment rather than based on long deliberation in situations that allow them do so, faster decision can even be associated with better performance. This conjecture also follows from the literature in psychology that has emphasized the role of intuition reflected in fast perception and pattern recognition \citep{deGroot1978,kahneman2003,kahnemanklein2009}.\footnote{The discussion about the existence of a speed-accuracy trade-off goes back to work by \cite{Henmon1911}. Work by \cite{Forster2003} has provided some tentative evidence that faster decision can also be associated with better performance in an experiment in a non-strategic environment using proof-reading as a decision task.}

Table \ref{tab:interaction_time_spent_deviate} reports results for an extended specification for the effects of subjective factors driving behavioral deviations that also accounts for the time spent on a move as an additional control variable. The coefficients for the subjective factors are qualitatively similar to the main results in Table \ref{tab:human_deviate} and seem to be unaffected by including the decision time spent on a move. The results also document, however, that spending more time on a move is associated with more frequent deviations from the benchmark and worse performance.

\begin{table}[!ht]
\begin{center}
\caption{Accounting for Decision Times \label{tab:interaction_time_spent_deviate}}
\footnotesize
{
\def\sym#1{\ifmmode^{#1}\else\(^{#1}\)\fi}
\begin{tabular*}{1\hsize}{@{\hskip\tabcolsep\extracolsep{-10pt}}l*{4}{D{.}{.}{-1}}}
\toprule
 &\multicolumn{4}{c}{Dependent Variable:}\\
                    &\multicolumn{1}{c}{$\Delta^E$ (binary)}&\multicolumn{1}{c}{$\Delta^P$ (binary)}&\multicolumn{1}{c}{$\Delta^N$ (binary)}&\multicolumn{1}{c}{$\Delta^C$ (categ.)}\\\cmidrule(lr){2-2}\cmidrule(lr){3-3}\cmidrule(lr){4-4}\cmidrule(lr){5-5}
                    &\multicolumn{1}{c}{(1)}         &\multicolumn{1}{c}{(2)}         &\multicolumn{1}{c}{(3)}         &\multicolumn{1}{c}{(4)}         \\
\midrule
\emph{Current Position}&                     &                     &                     &                     \\
Better position (>0.5 pawnunits)&      0.0690\sym{***}&     -0.0028         &      0.0719\sym{***}&     -0.0747\sym{***}\\
                    &    (0.0060)         &    (0.0043)         &    (0.0048)         &    (0.0069)         \\
Worse position (<-0.5 pawnunits)&      0.0108\sym{*}  &      0.0322\sym{***}&     -0.0214\sym{***}&      0.0536\sym{***}\\
                    &    (0.0060)         &    (0.0046)         &    (0.0050)         &    (0.0074)         \\
\bottomrule
\end{tabular*}
}
\end{center}
\footnotesize{\emph{Note:}{ OLS estimates. Evaluations of performance are based on the \textsc{Stockfish 8} chess engine. The variable \textit{Num. previous moves} is calculated as the number of previous moves per player. Standard errors are clustered on the game level. $^{*}$: $p<0.1$, $^{**}$: $p<0.05$, $^{***}$: $p<0.01$.}}

\end{table}


Additional results reveal that the time spent on a move also interacts with the psychological factors in determining behavioral deviations from the  benchmark (see Appendix Table \ref{tab:interaction_time_spent2}). In particular, more deliberation time in terms of time spent on a move tends to counteract the influence of being in a better or worse position, or of time pressure (in terms of less remaining time) on the likelihood of deviating from the benchmark. In terms of performance consequences, spending more time amplifies the positive performance consequences of being in a worse position and the negative performance consequences of time pressure. Moreover, longer deliberation time on a move tends to amplify the negative performance consequences associated with deviations from the benchmark due to fatigue by inducing more errors.

Together, these results indicate that faster decisions are associated with more behavioral deviations from the benchmark of cognitively bounded rationality and, at the same time, better performance than stipulated by the benchmark. In light of recent evidence that shorter deliberation times are associated with lower depth of reasoning \citep{alosferrer2020},
this suggests a superior intuitive assessment of positional configurations by humans, which is presumably related to intuition and experience, and which is particularly pronounced during critical phases of the game. These patterns are consistent with predictions of models of salience and selective memory \citep{Gennaioli/Shleifer:2010,Bordalo/Gennaioli/Shleifer:2020} under the assumption that chess experts have a very quick and intuitive perception of the best continuation and of critical positions that require longer deliberation times. This is also consistent with a two-step approach where decisions are taken either according to rational considerations or intuitively on a case-by-case assessment \citep{Sahm/Weizsaecker:2016}. 

%
%
%


\section{Concluding Remarks}\label{sec: conclusion}

In this paper, we provided a systematic analysis of human behavior under cognitively bounded rationality. In terms of methodology, we constructed a benchmark of cognitively bounded rationality that utilizes the artificial intelligence embedded in the algorithms of a chess engine that is subject to comparable computational limitations as humans. This allowed us to develop an identification strategy of the factors driving behavioral deviations from this objective benchmark as well as of the performance consequences of these deviations. This methodology might be useful for analyzing human behavior in other applications.

The empirical findings of this paper have important implications. The results show that professional chess players deviate systematically from the boundedly rational benchmark represented by a chess engine of comparable strength. In particular, the results indicate that time pressure, fatigue, complexity, and pressure from being in a better or worse position induce these deviations. However, the results also show that these deviations do not necessarily affect performance negatively, but often even entail superior performance. We also find that faster decisions are associated with more frequent deviations from the benchmark of bounded rationality and, concurrently, better performance. In light of previous theoretical literature and additional empirical results, this suggests that the superior performance is presumably due to experience or intuition of experts that provides them with a fairly fast and precise assessment of the decision problem and of the best choice.

While this paper contributes a new methodology to identify behavioral patterns of bounded rationality as well as its consequences for performance, the results are not conclusive about the underlying mechanisms. The results for decision times, behavioral deviations and performance suggest that first perception and recognition play an important role that needs to be understood better. Moreover, it is possible that behavioral deviations from the benchmark of cognitively bounded rationality are entirely due to mechanisms related to cognitive processes that underly the decisions of an individual player. It is equally possible, however, that the deviations are part of a strategy that incorporates beliefs about likely deviations of the opponent from the benchmark of boundedly rational behavior, thus incorporating the notion that behavior as stipulated by a chess engine might not be the optimal strategy, in analogy to the optimal strategy in a guessing game. Both possibilities are consistent with the empirical approach and the results presented in this paper. A natural next step in the research agenda is to apply the methodology developed here to investigate the respective behavioral mechanisms in more detail.

\newpage
\normalsize \renewcommand{\baselinestretch}{1}\normalsize

\bibliographystyle{ecta}
\bibliography{references}

\clearpage
\appendix 

\renewcommand{\thetable}{A\arabic{table}}
\setcounter{table}{0}

\renewcommand{\thefigure}{A\arabic{figure}}
\setcounter{figure}{0}

\section*{Appendix with Supplementary Material \newline For Online Publication}

\subsection*{Additional Figures}

%

\begin{figure}[h!]
\begin{center}
\includegraphics[width=0.45\textwidth]{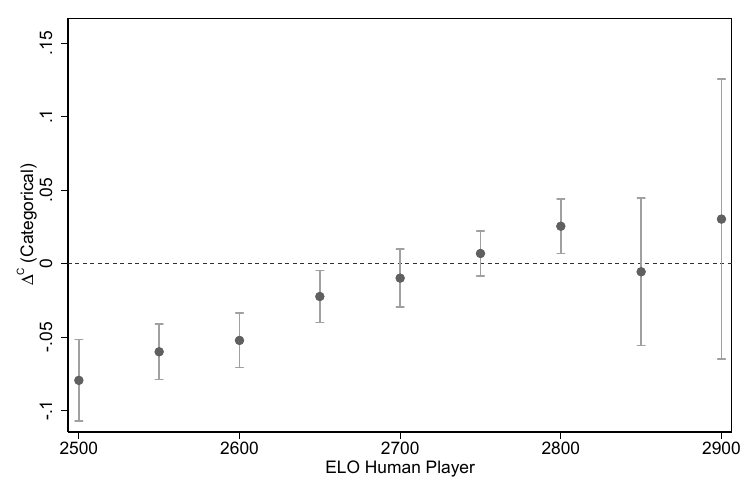}
\caption{Player Strength and Average Performance Difference between Human Players and Restricted Chess Engine \label{fig:Human_and_Restricted_Chess_Engine}}
\end{center}
\vspace{0.3cm}
\parbox{15cm}{\footnotesize \emph{Note:} This figure plots the difference of the performance of a player in comparison to the performance of the restricted chess engine. The graph is based on configurations in which human players had more than one hour remaining time budget and hence had no binding time constraints.  ELO numbers of players' depicted on the horizontal axis are split into equal-spaced intervals to compute the average within the interval. Whiskers report 95\% confidence intervals.}
\end{figure}

\clearpage
\newpage

\subsection*{Additional Tables}

\begin{table}[ht!]
\begin{center}
\caption{List of Tournaments in Dataset\label{tab:tournaments}}
\footnotesize
   \begin{tabular}{lrr}
	\hline \hline
 & Games & Percent 	\\
 \hline
\multicolumn{3}{c}{National Championships} \\
\hline
Russian Championship&165&8.3\%  \\
U.S. Championship&114 &5.8\% \\
Ukrainian Championship&78 &3.9\% \\
French Championship&62 &3.1\% \\
Armenian Championship&60 &3.0\% \\
Other National Championships&115&5.8\%  \\

\hline
\multicolumn{3}{c}{Invited Tournaments} \\
\hline
Wijk aan Zee&450&22.7\%  \\
Norway Chess&84&4.2\%  \\
Poikovsky&81&4.1\%  \\
Shamkir&76&3.8\%  \\
Lake Sevan&69&3.5\%  \\
Danzhou&42&2.1\%  \\
Other Invited Tournaments&306&15.4\%  \\

\hline
\multicolumn{3}{c}{World Chess Federation Tournaments} \\
\hline
 FIDE Grand Prix&280&14.1\%  \\
\hline
Sum&1,982&100.0\%   \\ \hline \hline
    \end{tabular}%
\end{center}
\end{table}

\begin{table}[ht!]
\begin{center}
\caption{Descriptive Statistics -- Game Level\label{tab:summary_games}}
\footnotesize
{
\def\sym#1{\ifmmode^{#1}\else\(^{#1}\)\fi}
\begin{tabular}{l*{1}{ccccc}}
\toprule
                    &       Games&        Mean&   Std. Dev.&         Min&         Max\\
\midrule
\emph{Player strength}&            &            &            &            &            \\
Elo rating white player&       1,982&      2681.7&       85.24&        2500&        2881\\
Elo rating black player&       1,982&      2681.0&       85.09&        2500&        2881\\
\emph{Game result}  &            &            &            &            &            \\
White player wins   &       1,982&       0.279&       0.448&           0&           1\\
Draw                &       1,982&       0.566&       0.496&           0&           1\\
Black player wins   &       1,982&       0.156&       0.363&           0&           1\\
\emph{Duration}     &            &            &            &            &            \\
Num. moves overall  &       1,982&       41.05&       14.55&          15&          98\\
Duration game (hours) overall&       1,982&       3.332&       1.443&           0&       8.113\\
\bottomrule
\end{tabular}
}
\end{center}
\footnotesize{\emph{Note:} The variable \textit{Num. moves overall} is calculated as the number of moves per player in game.}
\end{table}

\begin{table}[h!]
	\caption{Descriptive Statistics --  Move Level\label{tab: descriptive statistics}}
	\footnotesize
{
\def\sym#1{\ifmmode^{#1}\else\(^{#1}\)\fi}
\begin{tabular}{l*{1}{ccccc}}
\toprule
                    &       Moves&        Mean&   Std. Dev.&         Min&         Max\\
\midrule
\emph{Game characteristics}&            &            &            &            &            \\
Elo rating player   &     106,391&      2678.6&       86.08&        2500&        2881\\
Elo difference between players&     106,391&      -0.103&       71.07&        -284&         284\\
\emph{Performance measures}&            &            &            &            &            \\
$\Delta^E$ (binary)    &     106,391&       0.398&       0.489&           0&           1\\
$\Delta^P$ (binary)     &     106,391&       0.175&       0.380&           0&           1\\
$\Delta^N$ (binary)      &     106,391&       0.223&       0.416&           0&           1\\
\bottomrule
\end{tabular}
}
\vspace{0.1cm}
\footnotesize{\emph{Note:} Descriptive statistics for the baseline sample. Evaluations of performance are based on the \textsc{Stockfish 8} chess engine. The variable \textit{Distance second best move} contains missing values for configurations where there is only one legal move available to the player. The variable \textit{Num. previous moves} is calculated as the number of previous moves per player. The variable \textit{Remaining time (opp.)} has missing values because the remaining time of the opponent is not recorded for the final move of a game.}
\end{table}

%

\begin{table}[ht!]
\caption{Robustness -- Specifications Without Player-Game Fixed Effects \label{tab:baseline_no_FE}}
\footnotesize
{
\def\sym#1{\ifmmode^{#1}\else\(^{#1}\)\fi}
\begin{tabular*}{1\hsize}{@{\hskip\tabcolsep\extracolsep{-14pt}}l*{4}{D{.}{.}{-1}}}
\toprule
 &\multicolumn{4}{c}{Dependent Variable:}\\
                    &\multicolumn{1}{c}{$\Delta^E$ (binary)}&\multicolumn{1}{c}{$\Delta^P$ (binary)}&\multicolumn{1}{c}{$\Delta^N$ (binary)}&\multicolumn{1}{c}{$\Delta^C$ (categ.)}\\\cmidrule(lr){2-2}\cmidrule(lr){3-3}\cmidrule(lr){4-4}\cmidrule(lr){5-5}
                    &\multicolumn{1}{c}{(1)}         &\multicolumn{1}{c}{(2)}         &\multicolumn{1}{c}{(3)}         &\multicolumn{1}{c}{(4)}         \\
\midrule
\emph{Game characteristics}&                     &                     &                     &                     \\
Elo player (divided by 100)&     -0.0090\sym{***}&      0.0055\sym{***}&     -0.0145\sym{***}&      0.0201\sym{***}\\
                    &    (0.0028)         &    (0.0019)         &    (0.0022)         &    (0.0030)         \\
Elo difference (divided by 100)&     -0.0077         &     -0.0012         &     -0.0064         &      0.0052         \\
                    &    (0.0063)         &    (0.0045)         &    (0.0049)         &    (0.0070)         \\
\bottomrule
\end{tabular*}
}
\footnotesize{\emph{Note:}{{ OLS estimates. Evaluations of performance are based on the \textsc{Stockfish 8} chess engine. The variable \textit{Num. previous moves} is calculated as the number of previous moves per player. Standard errors are clustered on the game level. $^{*}$: $p<0.1$, $^{**}$: $p<0.05$, $^{***}$: $p<0.01$.}}}
\end{table}

\begin{table}[ht!]
\caption{Robustness -- Specifications with Subjective Factors in Isolation \label{tab:baseline}}
\footnotesize
{
\def\sym#1{\ifmmode^{#1}\else\(^{#1}\)\fi}
\begin{tabular*}{1\hsize}{@{\hskip\tabcolsep\extracolsep{-24pt}}l*{5}{D{.}{.}{-1}}}
\toprule
 &\multicolumn{5}{c}{Dependent Variable:}\\
                    &\multicolumn{5}{c}{Performance: $\Delta^C$ (categ.)}                        \\\cmidrule(lr){2-6}
                    &\multicolumn{1}{c}{(1)}         &\multicolumn{1}{c}{(2)}         &\multicolumn{1}{c}{(3)}         &\multicolumn{1}{c}{(4)}         &\multicolumn{1}{c}{(5)}         \\
\midrule
\emph{Current Position}&                     &                     &                     &                     &                     \\
Better position (>0.5 pawnunits)&     -0.0951\sym{***}&                     &                     &                     &     -0.0784\sym{***}\\
                    &    (0.0068)         &                     &                     &                     &    (0.0069)         \\
Worse position (<-0.5 pawnunits)&      0.0336\sym{***}&                     &                     &                     &      0.0554\sym{***}\\
                    &    (0.0074)         &                     &                     &                     &    (0.0075)         \\
\bottomrule
\end{tabular*}
}
\footnotesize{\emph{Note:}{ OLS estimates. Evaluations of performance are based on the \textsc{Stockfish 8} chess engine. The variable \textit{Num. previous moves} is calculated as the number of previous moves per player. Standard errors are clustered on the game level. $^{*}$: $p<0.1$, $^{**}$: $p<0.05$, $^{***}$: $p<0.01$.
}}

\end{table}

\begin{table}[ht!]
\caption{Robustness -- Excluding Positions with Evaluation Equal to 0.00\label{tab:baseline_excluding_position_0}}
\footnotesize
{
\def\sym#1{\ifmmode^{#1}\else\(^{#1}\)\fi}
\begin{tabular*}{1\hsize}{@{\hskip\tabcolsep\extracolsep{-12pt}}l*{4}{D{.}{.}{-1}}}
\toprule
 &\multicolumn{4}{c}{Dependent Variable:}\\
                    &\multicolumn{1}{c}{$\Delta^E$ (binary)}&\multicolumn{1}{c}{$\Delta^P$ (binary)}&\multicolumn{1}{c}{$\Delta^N$ (binary)}&\multicolumn{1}{c}{$\Delta^C$ (categ.)}\\\cmidrule(lr){2-2}\cmidrule(lr){3-3}\cmidrule(lr){4-4}\cmidrule(lr){5-5}
                    &\multicolumn{1}{c}{(1)}         &\multicolumn{1}{c}{(2)}         &\multicolumn{1}{c}{(3)}         &\multicolumn{1}{c}{(4)}         \\
\midrule
\emph{Current Position}&                     &                     &                     &                     \\
Better position (>0.5 pawnunits)&      0.0223\sym{***}&     -0.0228\sym{***}&      0.0451\sym{***}&     -0.0679\sym{***}\\
                    &    (0.0062)         &    (0.0048)         &    (0.0052)         &    (0.0077)         \\
Worse position (<-0.5 pawnunits)&     -0.0522\sym{***}&      0.0064         &     -0.0586\sym{***}&      0.0650\sym{***}\\
                    &    (0.0066)         &    (0.0052)         &    (0.0055)         &    (0.0085)         \\
\bottomrule
\end{tabular*}
}
\footnotesize{\emph{Note:}{{ In this table, configurations that are evaluated with 0.00 by the chess engine due to a mutally beneficial move repetition are excluded. OLS estimates. Evaluations of performance are based on the \textsc{Stockfish 8} chess engine. The variable \textit{Num. previous moves} is calculated as the number of previous moves per player. Standard errors are clustered on the game level. $^{*}$: $p<0.1$, $^{**}$: $p<0.05$, $^{***}$: $p<0.01$.}}}
\end{table}

\begin{table}[ht!]
\begin{center}
\caption{Robustness -- Flexible Specifications of Relative Positional Standing \label{tab:Baseline_split_better_position_five}}
\footnotesize
{
\def\sym#1{\ifmmode^{#1}\else\(^{#1}\)\fi}
\begin{tabular*}{1\hsize}{@{\hskip\tabcolsep\extracolsep{-10pt}}l*{4}{D{.}{.}{-1}}}
\toprule
 &\multicolumn{4}{c}{Dependent Variable:}\\
                    &\multicolumn{1}{c}{$\Delta^E$ (binary)}&\multicolumn{1}{c}{$\Delta^P$ (binary)}&\multicolumn{1}{c}{$\Delta^N$ (binary)}&\multicolumn{1}{c}{$\Delta^C$ (categ.)}\\\cmidrule(lr){2-2}\cmidrule(lr){3-3}\cmidrule(lr){4-4}\cmidrule(lr){5-5}
                    &\multicolumn{1}{c}{(1)}         &\multicolumn{1}{c}{(2)}         &\multicolumn{1}{c}{(3)}         &\multicolumn{1}{c}{(4)}         \\
\midrule
1 Slight advantage (+/=)&      0.0909\sym{***}&      0.0159\sym{***}&      0.0750\sym{***}&     -0.0591\sym{***}\\
                    &    (0.0063)         &    (0.0049)         &    (0.0054)         &    (0.0082)         \\
2 Clear advantage (+/-)&      0.1028\sym{***}&      0.0134\sym{**} &      0.0893\sym{***}&     -0.0759\sym{***}\\
                    &    (0.0075)         &    (0.0053)         &    (0.0061)         &    (0.0087)         \\
3 Decisive advantage  (+-)&      0.1082\sym{***}&     -0.0027         &      0.1108\sym{***}&     -0.1135\sym{***}\\
\bottomrule
\end{tabular*}
}
\end{center}
\footnotesize{\emph{Note:}{{ OLS estimates. Evaluations of performance are based on the \textsc{Stockfish 8} chess engine. The variable \textit{Num. previous moves} is calculated as the number of previous moves per player. Advantages and disadvantages are calculated based on the usual chess conventions: A configuration that is evaluated by a chess engine as less than 0.3 pawn units better for one side is considered equal ($=$). A configuration that is evaluated as between 0.3 and 0.7 pawn units better for one side is considered as a slight advantage ($+/=$) or slight disadvantage ($=/-$), respectively. A configuration that is evaluated as between 0.7 and 1.6 pawn units better for one side is considered as a clear advantage ($+/-$) or clear disadvantage ($-/+$), respectively. Positions that are evaluated as 1.6 better for one side are considered as a decisive advantage ($+-$) or decisive disadvantage ($-+$), respectively. Standard errors are clustered on the game level. $^{*}$: $p<0.1$, $^{**}$: $p<0.05$, $^{***}$: $p<0.01$.}}}

\end{table}

\begin{landscape}
\begin{table}[ht!]
\begin{center}
\caption{Robustness -- Alternative Proxies as Explanatory Variables \label{tab:Baseline_five_alternate_measures}}
\footnotesize
{
\def\sym#1{\ifmmode^{#1}\else\(^{#1}\)\fi}
\begin{tabular*}{0.75\hsize}{@{\hskip\tabcolsep\extracolsep{-14pt}}l*{4}{D{.}{.}{-1}}}
\toprule
 &\multicolumn{4}{c}{Dependent Variable:}\\
                    &\multicolumn{1}{c}{$\Delta^E$ (binary)}&\multicolumn{1}{c}{$\Delta^P$ (binary)}&\multicolumn{1}{c}{$\Delta^N$ (binary)}&\multicolumn{1}{c}{$\Delta^C$ (categ.)}\\\cmidrule(lr){2-2}\cmidrule(lr){3-3}\cmidrule(lr){4-4}\cmidrule(lr){5-5}
                    &\multicolumn{1}{c}{(1)}         &\multicolumn{1}{c}{(2)}         &\multicolumn{1}{c}{(3)}         &\multicolumn{1}{c}{(4)}         \\
\midrule
Better Position (pawn units, continous)&      0.0002\sym{**} &     -0.0001         &      0.0003\sym{***}&     -0.0003\sym{***}\\
                    &    (0.0001)         &    (0.0001)         &    (0.0001)         &    (0.0001)         \\
Worse Position (pawn units, continous)&     -0.0013\sym{***}&     -0.0005\sym{***}&     -0.0009\sym{***}&      0.0004\sym{***}\\
                    &    (0.0001)         &    (0.0000)         &    (0.0001)         &    (0.0001)         \\
\emph{Time Pressure}&                     &                     &                     &                     \\
\bottomrule
\end{tabular*}
}
\end{center}
\footnotesize{\emph{Note:} OLS estimates. Evaluations of performance are based on the \textsc{Stockfish 8} chess engine. Better/worse position are measured in absolute pawn units. \emph{Less than 10 moves before first time control} is a dummy indicating that a player has less than 10 moves to play before reaching move 40 when additional time is added to each players' time budget. In contrast to the variable \emph{remaining time} in the main analysis, less than 10 moves before first time control implies more time pressure. Accordingly, the sign of the coefficients for time pressure variable is expected to be opposite of that in the baseline specification. \emph{Duration game} is the overall time that both players have already spent thinking about their moves. \emph{Distance second best move} measures how far in terms of pawn units the chess engine evaluates the current configuration to be worse in case the second best move is played compared to the best move. 658 observations are dropped compared to the baseline specification because there is no second legal move available to the player. Standard errors are clustered on the game level. $^{*}$: $p<0.1$, $^{**}$: $p<0.05$, $^{***}$: $p<0.01$.}
\end{table}
\end{landscape}

\begin{table}[ht!]
\begin{center}
\caption{Robustness -- Both Players with ELO Numbers Above 2000\label{tab:baseline above_2000_elo}}
\footnotesize
{
\def\sym#1{\ifmmode^{#1}\else\(^{#1}\)\fi}
\begin{tabular*}{1\hsize}{@{\hskip\tabcolsep\extracolsep{-10pt}}l*{4}{D{.}{.}{-1}}}
\toprule
 &\multicolumn{4}{c}{Dependent Variable:}\\
                    &\multicolumn{1}{c}{$\Delta^E$ (binary)}&\multicolumn{1}{c}{$\Delta^P$ (binary)}&\multicolumn{1}{c}{$\Delta^N$ (binary)}&\multicolumn{1}{c}{$\Delta^C$ (categ.)}\\\cmidrule(lr){2-2}\cmidrule(lr){3-3}\cmidrule(lr){4-4}\cmidrule(lr){5-5}
                    &\multicolumn{1}{c}{(1)}         &\multicolumn{1}{c}{(2)}         &\multicolumn{1}{c}{(3)}         &\multicolumn{1}{c}{(4)}         \\
\midrule
\emph{Current Position}&                     &                     &                     &                     \\
Better position (>0.5 pawnunits)&      0.0781\sym{***}&      0.0020         &      0.0761\sym{***}&     -0.0741\sym{***}\\
                    &    (0.0045)         &    (0.0031)         &    (0.0037)         &    (0.0052)         \\
Worse position (<-0.5 pawnunits)&      0.0077\sym{*}  &      0.0306\sym{***}&     -0.0229\sym{***}&      0.0535\sym{***}\\
                    &    (0.0045)         &    (0.0032)         &    (0.0038)         &    (0.0054)         \\
\bottomrule
\end{tabular*}
}
\end{center}
\footnotesize{\emph{Note:}{{ OLS estimates. Evaluations of performance are based on the \textsc{Stockfish 8} chess engine. The variable \textit{Num. previous moves} is calculated as the number of previous moves per player. Standard errors are clustered on the game level. $^{*}$: $p<0.1$, $^{**}$: $p<0.05$, $^{***}$: $p<0.01$.}}}
\end{table}

\begin{table}[ht!]
\caption{Robustness -- Both Players with ELO Numbers Between 2400 and 2600\label{tab:baseline btw_2400_2600_elo}}
\footnotesize
{
\def\sym#1{\ifmmode^{#1}\else\(^{#1}\)\fi}
\begin{tabular*}{1\hsize}{@{\hskip\tabcolsep\extracolsep{-10pt}}l*{4}{D{.}{.}{-1}}}
\toprule
 &\multicolumn{4}{c}{Dependent Variable:}\\
                    &\multicolumn{1}{c}{$\Delta^E$ (binary)}&\multicolumn{1}{c}{$\Delta^P$ (binary)}&\multicolumn{1}{c}{$\Delta^N$ (binary)}&\multicolumn{1}{c}{$\Delta^C$ (categ.)}\\\cmidrule(lr){2-2}\cmidrule(lr){3-3}\cmidrule(lr){4-4}\cmidrule(lr){5-5}
                    &\multicolumn{1}{c}{(1)}         &\multicolumn{1}{c}{(2)}         &\multicolumn{1}{c}{(3)}         &\multicolumn{1}{c}{(4)}         \\
\midrule
\emph{Current Position}&                     &                     &                     &                     \\
Better position (>0.5 pawnunits)&      0.0876\sym{***}&      0.0045         &      0.0830\sym{***}&     -0.0785\sym{***}\\
                    &    (0.0093)         &    (0.0067)         &    (0.0082)         &    (0.0116)         \\
Worse position (<-0.5 pawnunits)&      0.0128         &      0.0354\sym{***}&     -0.0226\sym{***}&      0.0580\sym{***}\\
                    &    (0.0088)         &    (0.0065)         &    (0.0074)         &    (0.0108)         \\
\bottomrule
\end{tabular*}
}
\footnotesize{\emph{Note:}{{ OLS estimates. Evaluations of performance are based on the \textsc{Stockfish 8} chess engine. The variable \textit{Num. previous moves} is calculated as the number of previous moves per player. Standard errors are clustered on the game level. $^{*}$: $p<0.1$, $^{**}$: $p<0.05$, $^{***}$: $p<0.01$.}}}
\end{table}

\begin{table}[ht!]
\caption{Robustness -- Both Players with ELO Numbers Between 2600 and 2800\label{tab:baseline btw_2600_2800_elo}}
\footnotesize
{
\def\sym#1{\ifmmode^{#1}\else\(^{#1}\)\fi}
\begin{tabular*}{1\hsize}{@{\hskip\tabcolsep\extracolsep{-10pt}}l*{4}{D{.}{.}{-1}}}
\toprule
 &\multicolumn{4}{c}{Dependent Variable:}\\
                    &\multicolumn{1}{c}{$\Delta^E$ (binary)}&\multicolumn{1}{c}{$\Delta^P$ (binary)}&\multicolumn{1}{c}{$\Delta^N$ (binary)}&\multicolumn{1}{c}{$\Delta^C$ (categ.)}\\\cmidrule(lr){2-2}\cmidrule(lr){3-3}\cmidrule(lr){4-4}\cmidrule(lr){5-5}
                    &\multicolumn{1}{c}{(1)}         &\multicolumn{1}{c}{(2)}         &\multicolumn{1}{c}{(3)}         &\multicolumn{1}{c}{(4)}         \\
\midrule
\emph{Current Position}&                     &                     &                     &                     \\
Better position (>0.5 pawnunits)&      0.0813\sym{***}&      0.0034         &      0.0779\sym{***}&     -0.0746\sym{***}\\
                    &    (0.0080)         &    (0.0059)         &    (0.0061)         &    (0.0090)         \\
Worse position (<-0.5 pawnunits)&      0.0069         &      0.0319\sym{***}&     -0.0250\sym{***}&      0.0569\sym{***}\\
                    &    (0.0081)         &    (0.0061)         &    (0.0065)         &    (0.0097)         \\
\bottomrule
\end{tabular*}
}
\footnotesize{\emph{Note:}{{ OLS estimates. Evaluations of performance are based on the \textsc{Stockfish 8} chess engine. The variable \textit{Num. previous moves} is calculated as the number of previous moves per player. Standard errors are clustered on the game level. $^{*}$: $p<0.1$, $^{**}$: $p<0.05$, $^{***}$: $p<0.01$.}}}
\end{table}

\begin{table}[t]
\caption{Robustness -- Rational Benchmark Constructed with \textsc{Komodo}-Engine \label{tab:human_deviate Komodo}}
\footnotesize
{
\def\sym#1{\ifmmode^{#1}\else\(^{#1}\)\fi}
\begin{tabular*}{1\hsize}{@{\hskip\tabcolsep\extracolsep{-10pt}}l*{4}{D{.}{.}{-1}}}
\toprule
 &\multicolumn{4}{c}{Dependent Variable:}\\
                    &\multicolumn{1}{c}{$\Delta^E$ (binary)}&\multicolumn{1}{c}{$\Delta^P$ (binary)}&\multicolumn{1}{c}{$\Delta^N$ (binary)}&\multicolumn{1}{c}{$\Delta^C$ (categ.)}\\\cmidrule(lr){2-2}\cmidrule(lr){3-3}\cmidrule(lr){4-4}\cmidrule(lr){5-5}
                    &\multicolumn{1}{c}{(1)}         &\multicolumn{1}{c}{(2)}         &\multicolumn{1}{c}{(3)}         &\multicolumn{1}{c}{(4)}         \\
\midrule
\emph{Current Position}&                     &                     &                     &                     \\
Better position (>0.5 pawnunits)&      0.0596\sym{***}&     -0.0034         &      0.0630\sym{***}&     -0.0664\sym{***}\\
                    &    (0.0053)         &    (0.0036)         &    (0.0042)         &    (0.0058)         \\
Worse position (<-0.5 pawnunits)&     -0.0009         &      0.0204\sym{***}&     -0.0213\sym{***}&      0.0417\sym{***}\\
                    &    (0.0054)         &    (0.0037)         &    (0.0045)         &    (0.0061)         \\
\bottomrule
\end{tabular*}
}
\footnotesize{\emph{Note:}{ OLS estimates. 
Performance of the restricted engine is computed using the \textsc{Komodo} chess engine. The variable \textit{Num. previous moves} is calculated as the number of previous moves per player. Standard errors are clustered on the game level. $^{*}$: $p<0.1$, $^{**}$: $p<0.05$, $^{***}$: $p<0.01$. }}
\end{table}



\begin{table}[h!]
\begin{center}
	\caption{Total Effect -- Results for Semi-Continuous Performance Measure (Log-Modulus)\label{tab:baseline log modulus}}
	\footnotesize
{
\def\sym#1{\ifmmode^{#1}\else\(^{#1}\)\fi}
\begin{tabular*}{0.9\hsize}{@{\hskip\tabcolsep\extracolsep{-20pt}}l*{4}{D{.}{.}{-1}}}
\toprule
 &\multicolumn{4}{c}{Dependent Variable:}\\
                    &\multicolumn{4}{c}{Performance: $\Delta^L$ (log-mod.)}  \\\cmidrule(lr){2-5}
\ &\multicolumn{2}{c}{\textsc{Stockfish}} 
&\multicolumn{2}{c}{\textsc{Komodo}} \\ 
&\multicolumn{1}{c}{Total}                     
&\multicolumn{1}{c}{Intensive}
&\multicolumn{1}{c}{Total}                     
&\multicolumn{1}{c}{Intensive} \\                     
&\multicolumn{1}{c}{Effect}                     
&\multicolumn{1}{c}{Margin}
&\multicolumn{1}{c}{Effect}                     
&\multicolumn{1}{c}{Margin} \\
\cmidrule(lr){2-5}
                    &\multicolumn{1}{c}{(1)}         &\multicolumn{1}{c}{(2)}         &\multicolumn{1}{c}{(3)}         &\multicolumn{1}{c}{(4)}         \\
\midrule
\emph{Current Position}&                     &                     &                     &                     \\
Better position (>0.5 pawnunits)&     -0.0472\sym{***}&     -0.0935\sym{***}&     -0.0582\sym{***}&     -0.1538\sym{***}\\
\bottomrule
\end{tabular*}
}
\end{center}
\footnotesize{\emph{Note:}{ OLS estimates. The dependent variable is the difference in the evaluation of a move relative to the best possible move in pawn units in terms of a log modulus transformation, such that $\Delta^L_{gic} = \sgn(\Delta_{gic})\cdot \ln(|\Delta_{gic}| + 1)$ with $\Delta_{gic}=P_{gic}-P^r_c$. Evaluations of performance are based on comparisons to the \textsc{Stockfish 8} chess engine as the super engine. In columns (1)-(2) performance of the restricted engine is computed using \textsc{Stockfish 8}. In columns (3)-(4) performance of the restricted engine is computed using the \textsc{Komodo} chess engine. The variable \textit{Num. previous moves} is calculated as the number of previous moves per player. Standard errors are clustered on the game level. $^{*}$: $p<0.1$, $^{**}$: $p<0.05$, $^{***}$: $p<0.01$. }}
\end{table}

\begin{table}[ht!]
\begin{center}
\caption{Behavioral Heterogeneity -- Interactions with Time Pressure \label{tab:interaction_time_pressure}}
\footnotesize
{
\def\sym#1{\ifmmode^{#1}\else\(^{#1}\)\fi}
\begin{tabular*}{1\hsize}{@{\hskip\tabcolsep\extracolsep{-24pt}}l*{4}{D{.}{.}{-1}}}
\toprule
 &\multicolumn{4}{c}{Dependent Variable:}\\
                    &\multicolumn{1}{c}{$\Delta^E$ (binary)}&\multicolumn{1}{c}{$\Delta^P$ (binary)}&\multicolumn{1}{c}{$\Delta^N$ (binary)}&\multicolumn{1}{c}{$\Delta^C$ (categ.)}\\\cmidrule(lr){2-2}\cmidrule(lr){3-3}\cmidrule(lr){4-4}\cmidrule(lr){5-5}
                    &\multicolumn{1}{c}{(1)}         &\multicolumn{1}{c}{(2)}         &\multicolumn{1}{c}{(3)}         &\multicolumn{1}{c}{(4)}         \\
\midrule
\emph{Current Position}&                     &                     &                     &                     \\
Better position (>0.5 pawnunits)&      0.1096\sym{***}&      0.0150\sym{**} &      0.0947\sym{***}&     -0.0797\sym{***}\\
                    &    (0.0091)         &    (0.0063)         &    (0.0075)         &    (0.0104)         \\
Worse position (<-0.5 pawnunits)&      0.0233\sym{***}&      0.0368\sym{***}&     -0.0135\sym{*}  &      0.0503\sym{***}\\
                    &    (0.0088)         &    (0.0065)         &    (0.0073)         &    (0.0108)         \\
\bottomrule
\end{tabular*}
}
\end{center}
\footnotesize{\emph{Note:}{{ OLS estimates. Evaluations of performance are based on the \textsc{Stockfish 8} chess engine. The variable \textit{Num. previous moves} is calculated as the number of previous moves per player. Standard errors are clustered on the game level. $^{*}$: $p<0.1$, $^{**}$: $p<0.05$, $^{***}$: $p<0.01$.}}}
\end{table}

\begin{landscape}
\begin{table}[h!]
\begin{center}
\caption{Behavioral Heterogeneity -- The Role of Player Strength \label{tab:effect heterogeneity}}
\footnotesize
{
\def\sym#1{\ifmmode^{#1}\else\(^{#1}\)\fi}
\begin{tabular*}{0.8\hsize}{@{\hskip\tabcolsep\extracolsep{-20pt}}l*{4}{D{.}{.}{-1}}}
\toprule
 &\multicolumn{4}{c}{Dependent Variable:}\\
                    &\multicolumn{1}{c}{$\Delta^E$ (binary)}&\multicolumn{1}{c}{$\Delta^P$ (binary)}&\multicolumn{1}{c}{$\Delta^N$ (binary)}&\multicolumn{1}{c}{$\Delta^C$ (categ.)}\\\cmidrule(lr){2-2}\cmidrule(lr){3-3}\cmidrule(lr){4-4}\cmidrule(lr){5-5}
                    &\multicolumn{1}{c}{(1)}         &\multicolumn{1}{c}{(2)}         &\multicolumn{1}{c}{(3)}         &\multicolumn{1}{c}{(4)}         \\
\midrule
\emph{Current Position}&                     &                     &                     &                     \\
Better position (>0.5 pawnunits)&      0.0775\sym{***}&     -0.0044         &      0.0819\sym{***}&     -0.0863\sym{***}\\
                    &    (0.0081)         &    (0.0055)         &    (0.0068)         &    (0.0093)         \\
Worse position (<-0.5 pawnunits)&     -0.0010         &      0.0267\sym{***}&     -0.0278\sym{***}&      0.0545\sym{***}\\
                    &    (0.0077)         &    (0.0058)         &    (0.0062)         &    (0.0093)         \\
\bottomrule
\end{tabular*}
}
\end{center}
\footnotesize{\emph{Note:}{{ OLS estimates. Evaluations of performance are based on the \textsc{Stockfish 8} chess engine. The variable \textit{Num. previous moves} is calculated as the number of previous moves per player. Standard errors are clustered on the game level. $^{*}$: $p<0.1$, $^{**}$: $p<0.05$, $^{***}$: $p<0.01$.}}}
\end{table}
\end{landscape}

\begin{table}[h!]
\caption{Behavioral Heterogeneity -- Accounting for Color and Favorite Status \label{tab:RefDep_Better_Worse}}
\footnotesize
{
\def\sym#1{\ifmmode^{#1}\else\(^{#1}\)\fi}
\begin{tabular*}{1\hsize}{@{\hskip\tabcolsep\extracolsep{-14pt}}l*{4}{D{.}{.}{-1}}}
\toprule
 &\multicolumn{4}{c}{Dependent Variable:}\\
                    &\multicolumn{1}{c}{$\Delta^E$ (binary)}&\multicolumn{1}{c}{$\Delta^P$ (binary)}&\multicolumn{1}{c}{$\Delta^N$ (binary)}&\multicolumn{1}{c}{$\Delta^C$ (categ.)}\\\cmidrule(lr){2-2}\cmidrule(lr){3-3}\cmidrule(lr){4-4}\cmidrule(lr){5-5}
                    &\multicolumn{1}{c}{(1)}         &\multicolumn{1}{c}{(2)}         &\multicolumn{1}{c}{(3)}         &\multicolumn{1}{c}{(4)}         \\
\midrule
\emph{Current Position}&                     &                     &                     &                     \\
Better position (>0.5 pawnunits)&      0.0966\sym{***}&      0.0031         &      0.0935\sym{***}&     -0.0904\sym{***}\\
                    &    (0.0151)         &    (0.0115)         &    (0.0123)         &    (0.0185)         \\
Worse position (<-0.5 pawnunits)&      0.0032         &      0.0289\sym{***}&     -0.0256\sym{***}&      0.0545\sym{***}\\
                    &    (0.0098)         &    (0.0076)         &    (0.0080)         &    (0.0121)         \\
\bottomrule
\end{tabular*}
}
\footnotesize{\emph{Note:}{ OLS estimates. Evaluations of performance are based on the \textsc{Stockfish 8} chess engine. The variable \textit{White player} is a dummy variable indicating the player that plays with white pieces.
The variable \textit{Favorite} is a dummy variable indicating the player with the higher ELO number prior to the game.
The variable \textit{Num. previous moves} is calculated as the number of previous moves per player. Standard errors are clustered on the game level. $^{*}$: $p<0.1$, $^{**}$: $p<0.05$, $^{***}$: $p<0.01$.}}
\end{table}

\begin{table}[h!]
\begin{center}
\caption{Behavioral Heterogeneity -- Opponent's Remaining Time and Time Spent\label{tab:time_spent by opponent}}
\footnotesize
{
\def\sym#1{\ifmmode^{#1}\else\(^{#1}\)\fi}
\begin{tabular*}{0.95\hsize}{@{\hskip\tabcolsep\extracolsep{-15pt}}l*{4}{D{.}{.}{-1}}}
\toprule
 &\multicolumn{4}{c}{Dependent Variable:}\\
                    &\multicolumn{1}{c}{$\Delta^E$ (binary)}&\multicolumn{1}{c}{$\Delta^P$ (binary)}&\multicolumn{1}{c}{$\Delta^N$ (binary)}&\multicolumn{1}{c}{$\Delta^C$ (categ.)}\\\cmidrule(lr){2-2}\cmidrule(lr){3-3}\cmidrule(lr){4-4}\cmidrule(lr){5-5}
                    &\multicolumn{1}{c}{(1)}         &\multicolumn{1}{c}{(2)}         &\multicolumn{1}{c}{(3)}         &\multicolumn{1}{c}{(4)}         \\
\midrule
\emph{Current Position}&                     &                     &                     &                     \\
Better position (>0.5 pawnunits)&      0.0792\sym{***}&      0.0005         &      0.0787\sym{***}&     -0.0782\sym{***}\\
                    &    (0.0060)         &    (0.0043)         &    (0.0048)         &    (0.0069)         \\
Worse position (<-0.5 pawnunits)&      0.0015         &      0.0304\sym{***}&     -0.0289\sym{***}&      0.0594\sym{***}\\
                    &    (0.0061)         &    (0.0047)         &    (0.0051)         &    (0.0076)         \\
\bottomrule
\end{tabular*}
}
\end{center}
\footnotesize{\emph{Note:}{ OLS estimates. Evaluations of performance are based on the \textsc{Stockfish 8} chess engine. The variable \textit{Num. previous moves} is calculated as the number of previous moves per player. Standard errors are clustered on the game level. $^{*}$: $p<0.1$, $^{**}$: $p<0.05$, $^{***}$: $p<0.01$.}
}
\end{table}

\begin{landscape}
\begin{table}[ht!]
\footnotesize
\begin{center}
\caption{Time Spent Per Move -- Interactions  \label{tab:interaction_time_spent2}}
{
\def\sym#1{\ifmmode^{#1}\else\(^{#1}\)\fi}
\begin{tabular*}{0.8\hsize}{@{\hskip\tabcolsep\extracolsep{-18pt}}l*{4}{D{.}{.}{-1}}}
\toprule
 &\multicolumn{4}{c}{Dependent Variable:}\\
                    &\multicolumn{1}{c}{$\Delta^E$ (binary)}&\multicolumn{1}{c}{$\Delta^P$ (binary)}&\multicolumn{1}{c}{$\Delta^N$ (binary)}&\multicolumn{1}{c}{$\Delta^C$ (categ.)}\\\cmidrule(lr){2-2}\cmidrule(lr){3-3}\cmidrule(lr){4-4}\cmidrule(lr){5-5}
                    &\multicolumn{1}{c}{(1)}         &\multicolumn{1}{c}{(2)}         &\multicolumn{1}{c}{(3)}         &\multicolumn{1}{c}{(4)}         \\
\midrule
Time spent on move (min.)&      0.0390\sym{***}&      0.0149\sym{***}&      0.0241\sym{***}&     -0.0092\sym{***}\\
                    &    (0.0022)         &    (0.0018)         &    (0.0020)         &    (0.0030)         \\
\emph{Current Position}&                     &                     &                     &                     \\
Better position (>0.5 pawnunits)&      0.0780\sym{***}&     -0.0009         &      0.0789\sym{***}&     -0.0798\sym{***}\\
                    &    (0.0066)         &    (0.0048)         &    (0.0052)         &    (0.0075)         \\
\bottomrule
\end{tabular*}
}
\end{center}
\footnotesize{\emph{Note:}{ OLS estimates. Evaluations of performance are based on the \textsc{Stockfish 8} chess engine. The variable \textit{Num. previous moves} is calculated as the number of previous moves per player. Standard errors are clustered on the game level. $^{*}$: $p<0.1$, $^{**}$: $p<0.05$, $^{***}$: $p<0.01$.}}
\end{table}
\end{landscape}

\clearpage
\end{document}